\documentclass[10pt]{article}
\usepackage{amsmath,amssymb,amsthm,cite,epsfig}
\usepackage{graphicx}
\usepackage{caption}
\usepackage{subcaption}

\usepackage[usenames, dvipsnames]{color}
\usepackage{hyperref} 

\usepackage{authblk}
\usepackage[toc,page]{appendix}
\numberwithin{equation}{section}

\def \beq{\begin{equation}}
\def \eeq{\end{equation}}
\def \beqa{\begin{eqnarray}}
\def \eeqa{\end{eqnarray}}

\def \e{\varepsilon}
\def \l{\left(}
\def \r{\right)}

\begin{document}
\title{\begin{flushright}
\end{flushright}
{\bf Equilibrium properties of blackbody radiation with an ultraviolet energy
cut-off}}
\author[2]{Dheeraj Kumar Mishra\thanks{dkmishra@imsc.res.in}}
\author[1]{Nitin Chandra\thanks{nitin.c.25@gmail.com}}
\author[2]{Vinay Vaibhav\thanks{vinayv@imsc.res.in}}
\affil[1]{B1(102), Om Sai Enclave, Hirabagh, Hazaribagh, 825301, India}
\affil[2]{The Institute of Mathematical Sciences, Chennai, Tamil Nadu, 600113, 
India.}
\affil[2]{Homi Bhabha National Institute,
Training School Complex, Anushakti Nagar, Mumbai, 400085, India}
\date{\empty}
\maketitle

\begin{abstract}

We study various equilibrium thermodynamic properties of blackbody radiation
(i.e. a photon gas) with an ultraviolet energy cut-off.
We find that the energy density, specific heat etc. follow usual acoustic phonon
dynamics as have been well studied by Debye. Other thermodynamic quantities like
pressure, entropy etc. have also been calculated. 
The usual Stefan-Boltzmann law gets modified. 
We observe that the values of the thermodynamic quantities with the energy
cut-off is lower than the corresponding values in the theory without any such
scale.
The phase-space measure is also expected to get modified for an exotic spacetime
appearing at Planck scale, 
which in turn leads to the modification of Planck energy density distribution
and the Wien's displacement law. We found that the non-perturbative nature of
the thermodynamic quantities in the SR limit (for both the case with
ultravilolet cut-off and the modified
measure case), due to nonanalyticity
of the leading term, is a general feature of the theory accompanied with an
ultraviolet energy cut-off. We have also discussed the possible modification in
the case of Big Bang 
and the Stellar objects and have suggested a table top
experiment for verification in effective low energy case.

\end{abstract}

\section{Introduction}
We, in this article try to explore the modification in the known physics given
an ultraviolet cut-off in the theory.
It seems that in all the theories attempting to combine gravity with quantum
mechanics, a natural length/energy scale emerges, i.e. Planck length/energy.
This scale acts as a threshold where a new description of spacetime
is expected to appear. 
Doubly Special Relativity (DSR) attempts to incorporate
this threshold as an invariant quantity under a relativistic
transformation \cite{AmelinoCamelia:1997gz, AmelinoCamelia:2000mn,
AmelinoCamelia:2000ge}. 
The motivation of DSR theories is also derived from the observation of
interesting effects such as deformation of dispersion relation etc. at very high
energy
scales \cite{AmelinoCamelia:1997gz, AmelinoCamelia:1999pm, Jacobson:2005bg,
Shao:2009bv, Shao:2010wk}.  
The introduction of an observer independent energy scale in DSR formulation, say
$\kappa$, leads to such a modification in the dispersion relation of a free
particle \cite{AmelinoCamelia:1997gz, AmelinoCamelia:2000mn, Magueijo:2001cr,
Magueijo:2002am}. 
The energy threshold also acts as a cut-off on the highest possible energy value
in the physical (sub-Planckian)
world \cite{Magueijo:2002am,Chandra:2011nj}. 
In the formulation of DSR by Magueijo and Smolin (MS formalism)
\cite{Magueijo:2002am} this sub-Planckian regime
which is characterized by the energy $E \leq \kappa$ and the momentum $p \leq
\kappa$ is the result of the
choice of the $U-$map (this is a map between the standard Lorentz generators and
the modified ones resulting in the modification of the Poincare algebra keeping
the Lorentz sector intact). This choice of $U-$map is in sync with the
expectation of the emergence of the granular structure of spacetime at Planck
scale.
Similar cut-offs in momentum and/or energy are seen in other DSR formalisms as
well \cite{KowalskiGlikman:2002we}.
DSR formalism can also be extended to curved spacetime.
One such extension was proposed by MS and has been since then
studied from various
perspectives \cite{Magueijo:2002xx, Smolin:2005cz, Gorji:2016laj, Ali:2014zea,
Ling:2005bp}. DSR has also been explored from the point
of view of modified/deformed algebra called $\kappa-$Poincare
algebra \cite{KowalskiGlikman:2002we, KowalskiGlikman:2001ct, Blaut:2003wg,
Borowiec:2009vb, Magpantay:2010zz,Pramanik:2012fj}.
Also, we see similar cut-offs appearing in other candidate quantum gravity
theories like
noncommutative geometry, string theory, loop quantum gravity and GUP
(Generalized Uncertainty Principle)
etc. \cite{AmelinoCamelia:2000mn, Snyder:1946qz, Chandra:2014qva, Gross:1987ar,
Amati:1988tn, Maggiore:1993kv, Garay:1994en, Kempf:1994su, Kempf:1996nk,
Smailagic:2003yb, Smailagic:2003rp, Kober:2010um, Nozari:2012gd}.

Interestingly, as has been studied by many, it is possible to keep the Lorentz
group/algebra intact
for the DSR theories. On the other hand, the representation of the Lorentz
group/
algebra becomes non-linear to accommodate the invariant energy/length scale (for
example see \cite{Magueijo:2001cr, Magueijo:2002am} and the references
therein) as also stated above. Preserving the Lorentz group/algebra keeps the
theory simple and
intuitive. For our present study, we will follow the DSR formulation developed
in \cite{Magueijo:2001cr, Magueijo:2002am} by MS 
where the dispersion relation modifies to,

\begin{equation}
\e^2-p^2=m^2\left(1-\frac{\e}{\kappa}\right)^2
\label{MS}
\end{equation}
The Special Relativistic (SR) limit, i.e. $\kappa\rightarrow \infty$ gives the
usual dispersion relation,
\begin{equation}
\e^2-p^2=m^2
\end{equation}
We will stick to the natural units ($\hbar=1,c=1, k_B=1$) if not stated
explicitly. 

It is obvious that the modification in dispersion relation and the presence of
ultraviolet cut-off in energy introduced in the DSR theory will affect the
thermodynamics of many well studied
systems \cite{KowalskiGlikman:2001ct, Camacho:2007qy, Zhang:2011ms,
Grether:2007ur, Gregg:2008jb, Moussa:2015yqy, Chandra:2011nj, Das:2010gk}.
In \cite{Chandra:2011nj} an extensive study of classical ideal gas
thermodynamics has been done using MS formalism. On the other hand
\cite{Das:2010gk} studies the photon gas thermodynamics in the same DSR
formalism. 
It should be noted that the study in \cite{Das:2010gk} contains flaws.
They have considered the photon gas as a canonical ensemble obeying classical
(Maxwell-Boltzmann) statistics. On the other hand, it is a well known fact that
photon gas
follows a grand canonical ensemble (due to non-conservation of the photon
number) and obeys the quantum (Bose-Einstein) statistics. Because of this 
error in their formalism, the results obtained do not match with the usual
photon
gas thermodynamics (for example see section 7.3 of \cite{pathria}). 
Surprisingly, they match their results to the massless limit of the classical
ideal gas thermodynamics in SR.
For a photon, the mass being zero, dispersion relation remains same as in the
case of SR, i.e. $\e=p$. The DSR effect for the equilibrium properties of
blackbody radiation is basically
due to the ultraviolet cut-off $\kappa$ in energy. We model the blackbody
radiation in
equilibrium as a grand canonical ensemble of photons obeying Bose-Einstein
statistics as usually done\footnote{While the
draft of this paper was being prepared we came to know about a very recent
article by Mir Mehedi Faruk and Md. Muktadir Rahman \cite{Faruk:2016brl} 
where they have independently calculated the thermodynamic quantities of a
photon gas for the unmodified measure in $d$ dimension. Their article presents a
miscalculated result leading to the mistaken analysis and subsequently
misleading conclusions. 
This can be understood with the help of the following arguments:
\begin{enumerate}
\item As has been shown in this paper (see the discussion after (\ref{Zn}))
there is a one to one correspondence between photons in DSR and the usual
acoustic phonons. 
The specific heat in case of acoustic phonons has a constant high temperature
behaviour (which matches with the classical value given by the well known
Dulong-Petit law). 
On the other hand, one can easily notice the wrong result in
\cite{Faruk:2016brl}
according to which the specific heat goes as $T^3$ for all the temperature
values (see equation (40) therein with $d=3$).
 \item The point mentioned above is just one example. In fact, their expression
for the free energy itself is wrong leading to all the results of the paper
being wrong. Note that it is questionable to write 
the free energy in terms of Incomplete Gamma functions. On the other hand, the
free energy is related to the Incomplete Zeta functions as will become clear
later in our paper. 
Also, in calculating the free energy, they have removed a logarithmic term (the
boundary term in the integration) in a completely arbitrary manner. 
Note that this term is very important which will modify the expressions for
different thermodynamic quantities like entropy, pressure etc. in a significant
manner.
\end{enumerate}}.
We have also considered the most general possible modification in the phase
space measure for exotic spacetimes appearing at Planck scale. 
To list a few examples where similar modification appears, we note that in
noncommutative physics, which is one of the quantum gravity
candidate, the change in 
phase space appears as a change in the density of states as discussed in
\cite{Scholtz:2015fba}. Another candidate of the quantum gravity
theories, namely Loop Quantum Gravity also predicts the change in phase space
measure/density of
states at very high energies \cite{Corichi:2007tf, Hossain:2010wy,
Corichi:2012bg,
Majumder:2012qy}. We, in this article, being non specific will proceed with the
most generalized (isotropic and Taylor series expandable) modification in phase
space integration measure. 
In \cite{Chandra:2011nj}, a momentum dependent measure
has been considered which is a special case of our generalized approach. 
There have been many interesting attempts to incorporate the additivity of
energy and momentum of composite systems in
DSR \cite{Liberati:2004ju, Hossenfelder:2007fy, Hossenfelder:2014ifa,
Mandanici:2007eb, Deriglazov:2004hg, Girelli:2004ue, Deriglazov:2004yr}.
In this paper, we will not follow any particular prescription as the issue is
still not well settled.

The present paper starts with calculating various thermodynamic quantities such
as energy density, pressure,
entropy etc. with an ultraviolet energy cut-off. Next, we study the possible
changes due to the change in
phase space measure at Planck scale. We then go on calculating various
thermodynamic quantities
as energy density, pressure,
entropy etc. with such a modified measure. 
The possible realizations in case of Big Bang and Stellar objects have also been
discussed.
We have then analysed the low and
high temperature limits of all the thermodynamic quantities in both cases. 
Finally, we have also discussed the possible
physical realizations of the results obtained in effective low energy case. In
doing so we suggest a very
simple and intuitive table top experiment to test our results.
We, therefore, have discussed the DSR effects in three scenarios. The general
development with an ultraviolet energy cut-off is discussed in Section 2. The
modification at Planck scale, as a change in phase space measure, is discussed
in Section 3. The Planck scale effects
and effective Planck scale effects are studied in Section 4. And finally, the low
energy effective cut-off effects has been explored in Section 6.
We have also summarized the whole paper at the end. Some of the results are
listed in the appendix, in order not to break the continuity of the paper.

\section{Equilibrium properties of blackbody radiation with an ultraviolet
cut-off}\label{unmodified}
In this section, we will see the possible changes in thermodynamic quantities of
photon gas with an ultraviolet energy cut-off. The model
contains an ideal gas of identical and indistinguishable quanta namely,
photons,\cite{pathria}.
There are $n_\omega$ number of photons each with energy $\e=\omega$. The mean
value of $n_\omega$ is,  

\begin{align}\label{meanPhoton}
 \langle n_\omega \rangle=  \frac{ \displaystyle{\sum_{n_\omega=0}^{\infty}}
n_\omega e^{-\frac{n_\omega
\omega}{T}}}{\displaystyle{\sum_{n_\omega=0}^{\infty}} e^{-\frac{n_\omega
\omega}{T}}}=\frac{1}{e^{\frac{\omega}{T}}-1}
\end{align}
giving  mean energy as,

\begin{align}
 \langle \e \rangle=\omega \langle n_\omega
\rangle=\frac{\omega}{e^{\frac{\omega}{T}}-1}
\end{align}
In the large volume limit, the volume of the phase space can be used to find the
number of modes between the range $\omega$ and $\omega+d\omega$ which are given
by,

\begin{align}\label{densityPhoton}
 a(\omega) d\omega = \frac{2}{{(2 \pi)}^3} \left(4\pi p^2 dp \right) \int d^3x =
\frac{V_{ac} \omega^2 d\omega}{\pi^2}
\end{align}
Note that photons obey the dispersion relation $\omega=\e=p$. Factor $2$
comes due to the $2$ transverse polarizations of a photon. It is also to be
noted that the above expression will get modified when we consider the change of
the phase space measure in case of DSR.
The energy density distribution therefore becomes,

\begin{align}\label{distribution}
 u(\omega) d\omega= \frac{a(\omega) d\omega}{V_{ac}} \langle \e
\rangle=\frac{1}{\pi^2}\frac{\omega^3 d\omega}{e^{\frac{\omega}{T}}-1}
\end{align}
This is the usual Planck energy density distribution.
\subsection{Energy Density}\label{energySection}
Integrating (\ref{distribution}) from $\omega=0$ to $\omega=\kappa$  we get the
energy density of the photon gas as,

\begin{align}\label{energyDensity}
u \equiv \frac{U}{V_{ac}}=\int_0^{\kappa} u(\omega)d\omega =\frac{T^4}{\pi^2}
\int_0^{\frac{\kappa}{T}} \frac{x^3 dx}{e^x-1}=\frac{6
T^4}{\pi^2}\bigg[Z_4(0)-Z_4\left(\frac{\kappa}{T}\right)\bigg]
\end{align}
Here we have changed the variable to $x=\frac{\omega}{T}$.
Note that at finite and non-zero $T$, as $\kappa\rightarrow \infty$ this
expression reduces to the one given in $(12)$ on page $203$ in \cite{pathria},
giving usual law.
Also, $Z_n(x)$ is the incomplete zeta functions or ``Debye functions"(refer to
section $27.1$ of \cite{abramowitz}), and is given as,

\begin{align}\label{Zn}
 Z_n(x)=\frac{1}{\Gamma(n)}\int_x^{\infty} \frac{t^{n-1}}{e^t-1} dt
\end{align}
We note that $Z_n(0)=\zeta(n)$ where $\zeta(z)$ is the Riemann-Zeta function, in
particular $Z_4(0)=\zeta(4)=\frac{\pi^4}{90}$. It is remarkable that
(\ref{energyDensity}) is exactly same as in the case of acoustic
phonons\cite{ashcroft} with the replacements $\kappa \rightarrow \Theta_D$
(Debye temperature),
$2$ (number of photon polarizations) $\rightarrow$ $3$ (number of acoustic modes
in monoatomic Bravais lattice) and the velocity of acoustic phonons has to be
taken to be equal to $1$ for correct matching as we are working in natural
units. 
In case of acoustic phonons the cut-off on the possible frequencies comes due to
the finiteness of first Brillouin zone which itself is restricted by the number
density of ions in the lattice. On the other hand the energy cut-off $\kappa$
in (\ref{energyDensity}) comes from the quantum gravity considerations. We
expect the specific heat $C_V=\left(\frac{\partial U}{\partial
T}\right)_{V_{ac}}=T\left(\frac{\partial S}{\partial T}\right)_{V_{ac}}$ for a
photon gas with such an ultraviolet energy cut-off to follow the behaviour of
$C_V$ as in the case of acoustic
phonons.
For a mathematically rigorous
treatment of Debye theory see \cite{debye}.
Debye functions $Z_n(z)$ are related to the polylogarithm function $Li_n(z)$ by
(see (16.2) in \cite{polylogarithm})

\begin{align}\label{polyExpansion}
 Z_{n}(z)= \sum_{k=0}^{n-1} Li_{n-k}(e^{-z})\frac{z^k}{k!}
\end{align}
for $n>0$. Especially $Z_n(0)=Li_n(0)$. Here  polylogarithm functions themselves
can be series expanded for $|z|<1$ as (see (8.1) in \cite{polylogarithm})

\begin{align}
Li_n(z)=\displaystyle{\sum_{a=1}^{\infty}} \frac{z^a}{a^n}. 
\end{align}
The integral representation of $Li_n(z)$ is, for $Re(n) > 0$, as follows (see
(1) in \cite{polylogarithm})

\begin{align}
 Li_n(z)= \frac{z}{\Gamma(n)}\int_0^{\infty} \frac{t^{n-1}}{e^t-z}dt
\end{align}
In particular $Li_1(z)=-\ln(1-z)$ (see (6.1) in \cite{polylogarithm}). Thus the
energy density can be written in terms of $Li_n(z)$ as given below

\begin{align}\label{energyPoly}
 u = \frac{{\pi}^2
T^4}{15}-\bigg[\bigg(\frac{6T^4}{\pi^2}\bigg)Li_4\left(e^{-\frac{\kappa}{T}}
\right)+\bigg(\frac{6\kappa T^3}{\pi^2}\bigg)
Li_3\left(e^{-\frac{\kappa}{T}}\right)+\bigg(\frac{3{\kappa}^2 T^2}{\pi^2}\bigg)
 Li_2\left(e^{-\frac{\kappa}{T}}\right)+\bigg(\frac{{\kappa}^3 T}{\pi^2}\bigg)
Li_1\left(e^{-\frac{\kappa}{T}}\right) \bigg].
\end{align}
Note that the first term corresponds to the usual Stefan-Boltzmann law. All the
other terms modify the law which in turn, will give a correction to the
temperature measurements of different stellar objects. These correction terms
vanish in the SR
limit. Note that the SR limit $\frac{1}{\kappa} \rightarrow 0$ is nonanalytic in
nature and hence the energy density cannot be perturbatively expanded in a
Taylor series around this limit. This observation has also been seen in case of
classical ideal gas with an invariant energy scale \cite{Chandra:2011nj}. As the
only contribution for a photon gas is due to the ultraviolet cut-off introduced,
it is clear that the non-perturbative nature of the modified thermodynamics is a
consequence of this cut-off.
Also for all possible temperatures, the argument of the polylogarithm in
(\ref{energyPoly}) i.e. $e^{-\frac{\kappa}{T}}$ is a positive quantity making
(\ref{polyExpansion}) a positive number which leads to the correction term in
the expression of energy density being negative.   
This fact is clearly visible from the plot of energy density (see
figure~\ref{fig:ed} on page~\pageref{fig:ed}) where the plot with modified
energy density 
is always lower than the corresponding SR plot.
This fact can also be understood from the integral expression in
(\ref{energyDensity}) where the integrand is always a positive quantity and a
positive contribution $\frac{6 T^4}{\pi^2}
Z_4\left(\frac{\kappa}{T}\right)=\frac{T^4}{\pi^2}
\int_{\frac{\kappa}{T}}^{\infty} \frac{x^3 dx}{e^x-1}$ has been removed from the
SR value to get the corresponding modified value.  

\subsection{Specific heat}
We put $\frac{U}{T}=u\frac{V_{ac}}{T}$ and use (\ref{energyPoly}) along with
using the derivatives of polylogarithm given by (see (4.1) in
\cite{polylogarithm}) $\frac{\partial}{\partial
\mu}[Li_n(e^{\mu})]=Li_{n-1}(e^{\mu})$ and obtain the expression for specific
heat as,

\begin{align}\label{C_V}
 C_V =\bigg(\frac{\partial U}{\partial T}\bigg)_{V_{ac}} &=-\frac{{\kappa}^4
V_{ac}}{{\pi}^2 T} \frac{1}{(e^{\frac{\kappa}{T}}-1)}+ \bigg[\frac{4 {\pi}^2 T^3
V_{ac}}{15}-\left(\frac{24 T^3
V_{ac}}{\pi^2}\right)Li_4\left(e^{-\frac{\kappa}{T}}\right)-\left(\frac{24
\kappa T^2 V_{ac}}{\pi^2}\right) Li_3\left(e^{-\frac{\kappa}{T}}\right)\nonumber
\\
 &-\left(\frac{12{\kappa}^2 T V_{ac}}{\pi^2}\right) 
Li_2\left(e^{-\frac{\kappa}{T}}\right)- \frac{4 V_{ac} {\kappa}^3}{\pi^2}
Li_1\left(e^{-\frac{\kappa}{T}}\right)\bigg]= -\frac{{\kappa}^4 V_{ac}}{{\pi}^2
T} \frac{1}{(e^{\frac{\kappa}{T}}-1)}+4\bigg(\frac{U}{T}\bigg).
 \end{align}
As we have already seen that the energy density $u$ in the modified case has one
to one
correspondence with that of the acoustic phonon modes in Debye model, therefore
the specific heat $C_V =V_{ac}\bigg(\frac{\partial u}{\partial
T}\bigg)_{V_{ac}}=\left(\frac{T}{\kappa}\right)^3 \frac{\kappa^3 V_{ac}}{\pi^2}
\int_0^{\frac{\kappa}{T}} \frac{x^4 e^{x} dx}{(e^{x}-1)^2}$ will also follow the
same correspondence (see (17), (18), (19) etc. of section 7.4 in
\cite{pathria}).
Again the SR limit gives the usual result $C_V=4U/T$ ( note that $V_{ac}
\rightarrow V$ in this limit). The extra negative contribution in (\ref{C_V}) is
non-perturbative in the SR limit along with the non-perturbative contributions
from the term $4U/T$. Obviously, overall $C_V$ takes a lower value than the
corresponding SR values. This fact is visible from the plot also (see
figure~\ref{fig:sh} on page~\pageref{fig:sh}). The behaviour of $C_V$ for the
full range of $\frac{T}{\kappa} \in [0,1]$ is also shown in the
figure~\ref{fig:sh1} on page~\pageref{fig:sh1} which certainly mimics the Debye
theory. 
In the Debye theory, however, T may go up to infinity in which case the specific
heat goes to a constant value.

\subsection{Radiation Pressure}
The grand canonical partition function for the photon gas (with fugacity $z=1$)
is\cite{pathria} $Q(V_{ac},T)=\prod_\e \frac{1}{1-e^{-\frac{\e}{T}}}$ leading to
the expression for $q$-potential as $q\equiv\frac{PV_{ac}}{T}\equiv \ln
Q(V_{ac},T)=-\sum_\e \ln(1-e^{-\frac{\e}{T}})$.
In the large volume limit doing integration by parts we obtain,

\begin{align}\label{PressureUnmod}
P&=-\frac{T{\kappa}^3}{3{\pi}^2}
ln(1-e^{-\frac{\kappa}{T}})+\frac{T^4}{3{\pi}^2} \int_0^{\frac{\kappa}{T}}
\frac{x^3 dx}{e^{x}-1} \nonumber \\
&=-\frac{T{\kappa}^3}{3{\pi}^2} ln(1-e^{-\frac{\kappa}{T}})+ \frac{1}{3}u.
\end{align}
Thus the equation of state for the blackbody radiation field, i.e., the relation
between
the pressure and the energy density got modified and goes to the correct SR
limit $P_{SR}=\frac{1}{3}(u)_{SR}=\frac{{\pi}^2 (T_{SR})^4}{45}$.
The explicit temperature dependence of the radiation pressure is given by,

\begin{align}
 P = \frac{{\pi}^2
T^4}{45}-\bigg[\bigg(\frac{2T^4}{\pi^2}\bigg)Li_4\left(e^{-\frac{\kappa}{T}}
\right)+\bigg(\frac{2\kappa T^3}{\pi^2}\bigg)
Li_3\left(e^{-\frac{\kappa}{T}}\right)+\bigg(\frac{{\kappa}^2 T^2}{\pi^2}\bigg) 
Li_2\left(e^{-\frac{\kappa}{T}}\right)\bigg].
\end{align} 
The correction term in the above expression is negative (see figure~\ref{fig:rp}
on page~\pageref{fig:rp} ) and non-perturbative in nature (see discussion in
section \ref{energySection}). 

\subsection{Entropy}\label{sectionEntropy} 
The Helmholtz free energy is given by (the chemical potential $\mu=0$),

\begin{align}
 A=\mu N-PV_{ac}=-PV_{ac}= \frac{T{\kappa}^3V_{ac}}{3{\pi}^2}
ln(1-e^{-\frac{\kappa}{T}})-\bigg(\frac{U}{3}\bigg).
\end{align}
The entropy becomes,

\begin{align}
 S=\frac{U-A}{T}=-\frac{{\kappa}^3V_{ac}}{3{\pi}^2}
ln(1-e^{-\frac{\kappa}{T}})+\frac{4}{3} \bigg(\frac{U}{T}\bigg).
\end{align}
In SR limit, the first term vanishes and the expression goes to the correct
result $S_{SR}=\frac{4}{3}\left(\frac{U}{T}\right)_{SR}$ (see (19) of section
7.3 in \cite{pathria}).
As done in case of radiation pressure if we write the explicit temperature
dependence of the entropy, the negative non-perturbative contribution will be
very apparent which can be seen from the plot as well (see figure~\ref{fig:ent}
on page~\pageref{fig:ent}). The decrease in the entropy value for the modified
case
can be explained by the presence of ultraviolet cut-off which restricts the
number of available microstates to the system.
\subsection{Equilibrium number of photons}
The equilibrium number of photons can be obtained by integrating the product of
mean number of photons (\ref{meanPhoton}) and the volume of the phase space
(\ref{densityPhoton}),

\begin{align}
 \bar{N}= \int_0^{\kappa} \frac{V_{ac}}{\pi^2}\frac{\omega^2
d\omega}{e^{\frac{\omega}{T}}-1}=\frac{2 V_{ac}T^3}{{\pi}^2}
\left[Z_3(0)-Z_3\left(\frac{\kappa}{T}\right)\right]
\end{align}
Here $Z_3(0)=\zeta(3)$ is also called Apery's constant. This with the proper
replacements corresponds to the equilibrium number of acoustic phonons in the
Debye
theory. In the $\kappa \rightarrow \infty$ limit $Z_n(\frac{\kappa}{T})
\rightarrow 0$ 
and $V_{ac}\rightarrow V$ and we get the usual SR result
$(\bar{N})_{SR}=\frac{T^3V}{{\pi}^2}(2 \zeta(3))$ as given in (23) of section
7.3 in \cite{pathria}.
Like other thermodynamic quantities the equilibrium number of photons also gets
a negative non-perturbative correction.
The decrease in the $\bar{N}$ value for the modified case is also due to the
cut-off
which restricts the number of available normal modes.

\section{Photon gas thermodynamics at Planck scale for exotic
spacetimes}\label{phase}
In this section, we will discuss the possible modifications in the known
thermodynamic quantities if we consider a change in phase space measure along
with an invariant ultraviolet cut-off. 
Almost all the thermodynamic quantities for well studied systems encounter the
large volume limit where discrete summation over energy values goes to the
integration over phase space i.e. $\sum_{\e} \rightarrow \frac{1}{(2\pi)^3}\int
\int d^3x d^3p$.
But for exotic spacetimes appearing at Planck scale, we expect the phase-space
to modify (see for example \cite{Chandra:2011nj}) as $\sum_{\e} \rightarrow
\frac{1}{(2\pi)^3}\int \int d^3x  d^3p f(\vec x,\vec p)$.
Here we have considered the most general possible modification.
The thermodynamic quantities are derivable from the partition function of the
form $\frac{1}{\l2\pi\r^3} \int \int  d^3x d^3p  F(\e)$ which due to the change
in phase space measure modifies to $\frac{1}{\l2\pi\r^3} \int \int  d^3x d^3p
f(\vec x,\vec p) F(\e)$.
Assuming the spacetime to be isotropic and $f(\vec x,\vec p)= f(r,p)$ to be
Taylor series expandable in the powers of $\big(\frac{1}{r\kappa}\big)$ and
$\big(\frac{\e}{\kappa}\big)$ we get,

\begin{equation}
f(r,p) = \displaystyle{\sum_{n=0,n^{\prime}=0}^{\infty}}
\frac{a_{n,n^{\prime}}}{n!n^{\prime}!}
\left(\frac{\e}{\kappa}\right)^{n}\left(\frac{1}{r \kappa}\right)^{n^{\prime}},
\label{taylor}
\end{equation}
with $a_{0,0}=1$ as for $\kappa\rightarrow\infty$ we expect $f(r,p)\rightarrow
1$.
This expansion is valid only when $\frac{\e}{\kappa}$, $\frac{1}{r\kappa}<1$
throughout the integration range, this requires $\e<\kappa$ and
$r>\frac{1}{\kappa}$. Thus $\kappa$ acts as highest energy cut-off while
$\frac{1}{\kappa}$
acts as the lowest length cut-off.
Finally the integral changes to, 

\beqa
\frac{1}{\l2\pi\r^3} \displaystyle{\sum_{n=0,n^{\prime}=0}^{\infty}}
\frac{a_{n,n^{\prime}}}{n!n^{\prime}! \kappa^{n+n^{\prime}}}
\int_{r=\frac{1}{\kappa}}^{R} \int_{p=0}^{\kappa} d^3x d^3p\,\,  \e^{n}
\left(\frac{1}{r}\right)^{n^{\prime}} F(\e),
\eeqa
$R$ being the radius of the spherical volume considered.
Here we have interchanged the double summation and the integration which is
allowed if (see appendix \ref{App}),

\beqa
\displaystyle{\sum_{n=0,n^{\prime}=0}^{\infty}}
\frac{|a_{n,n^{\prime}}|}{n!n^{\prime}! \kappa^{n+n^{\prime}}}
\int_{r=\frac{1}{\kappa}}^{R} \int_{p=0}^{\kappa} d^3x d^3p\,\,  \e^{n}
\left(\frac{1}{r}\right)^{n^{\prime}} |F(\e)| < \infty.
\label{condition_modmeas_expansion}
\eeqa
Performing the integration over the coordinate space we obtain

\beqa
\frac{1}{\l2\pi\r^3} \int \int  d^3x d^3p f(r,p) F(\e) &=& \frac{1}{\l2\pi\r^3}
\displaystyle{\sum_{n=0,n^{\prime}=0 \atop n^{\prime}\neq 3}^{\infty}}
\frac{a_{n,n^{\prime}}}{n!n^{\prime}! \kappa^{n+3}} \frac{4\pi
}{\left(3-n^{\prime}\right)} \left[\left(\frac{3V\kappa^3}{4\pi}
\right)^{\frac{3-n^{\prime}}{3}}-1 \right]  \int_{p=0}^{\kappa} d^3p\,\,
\left({\e}\right)^{n} F(\e)\nonumber \\
&+& \frac{1}{\l2\pi\r^3}\displaystyle{\sum_{n=0}^{\infty}} \frac{a_{n,3}}{n!3!
\kappa^{n+3}} \left( \frac{4\pi}{3} \right) \ln\left(\frac{3V \kappa^3}{4 \pi}
\right) \int_{p=0}^{\kappa} d^3p\,\, \left({\e}\right)^{n} F(\e) 
\label{exotic}
\eeqa
where $V=\frac{4}{3}\pi R^3$ is the volume of the spherical ball of radius $R$.
The accessible part of the volume for the particle is $V_{ac}=V-\frac{4\pi}{3
\kappa^3}$. For the large volume limit the minimum length $\frac{1}{\kappa}<< R$
implying $V\kappa^3 >>1$
which in turn implies $V_{ac} \approx V$. Note that a small volume
$\frac{4\pi}{3 \kappa^3}$ is inaccessible to each particle. This inaccessible
volume can be extracted out at any point in the space as volume being large, all
the 
space points are equivalent. We have extracted out this volume at the centre
$r=0$.
$n'\geq 4$ as the powers of $\frac{1}{V\kappa^3}$ increases.

\subsection*{Example: Classical Ideal gas in canonical ensemble}
Let us take a particular example of $F(\e)$ to illustrate this further. We
consider the classical ideal gas in canonical ensemble obeying Maxwell-Boltzmann
statistics with the partition function \cite{pathria},

\begin{equation}
 Z_N\l V_{ac},T\r=\sum_E\exp[-\beta E]={1\over N!}[Z_1\l V_{ac},T\r]^N,
\end{equation}
where $Z_1(V_{ac},T)$ is the single particle partition function, $N$ is the
total number of constituent particles, $\beta={1\over T}$ and the total energy
$E$ of the system is $E=\sum_{\e}n_{\e}\e$.
Here $n_{\e}$ is the number of particles corresponding to the single particle
energy $\e$ and satisfies $\sum_{\e}n_{\e}=N$.
The single particle partition function is given by $Z_1\l
V_{ac},T\r=\sum_{\e}\exp[-\beta (\e-m_0)]$
In the large volume limit using (\ref{exotic}) for $F(\e)=\exp(-\beta(\e-m_0))$
and following the arguments given in \cite{Chandra:2011nj} we obtain,

\beqa
Z_1\l V_{ac},T\r &=& \displaystyle{\sum_{n=0,n^{\prime}=0 \atop n^{\prime}\neq
3}^{\infty}} \frac{a_{n,n^{\prime}}}{n!n^{\prime}! \kappa^{n}}
\left(\frac{3}{(3-n^{\prime})(\kappa^3 V_{ac})} \right)
\left[\left(\frac{3V\kappa^3}{4\pi} \right)^{\frac{3-n^{\prime}}{3}}-1 \right]
\,\, \l m_{0}-\frac{\partial}{\partial \beta}\r^{n} Z_1^0\l V_{ac},T\r 
\nonumber \\
&+& \displaystyle{\sum_{n=0}^{\infty}} \frac{a_{n,3}}{n!\kappa^{n}}
\left(\frac{4 \pi}{18 \kappa^3 V_{ac}} \right) \ln\left(\frac{3V\kappa^3}{4 \pi}
\right)  \,\, \l m_{0}-\frac{\partial}{\partial \beta}\r^{n} Z_1^0\l V_{ac},T\r 
\eeqa
where $Z_1^0\l V_{ac},T\r$ is the single particle partition function with the
unmodified measure,

\beqa
  Z_1^0 \l V_{ac},T\r = \frac{V_{ac}}{\l2\pi\r^3} \int_{p=0}^{\kappa} d^3p
\exp(-\beta(\e-m_0)).
\label{z10}
\eeqa
The expression for $Z_1\l V_{ac},T\r$ has now non-trivial dependence on $V$
unlike in the case of $Z_1^0\l V_{ac},T\r$. With this modification the value of
thermodynamic quantities, especially pressure, changes.
Let's not digress anymore and continue with the study of the photon gas
thermodynamics.

We will now consider the change in phase space as described above. This leads to
the modification of the energy density distribution as well as the $q$-potential
which in effect modifies all the thermodynamic quantities.

\subsubsection{Modified Planck's energy distribution and Wien's
law}\label{modWien}
It is clear that the change in phase space is going to modify the Planck
distribution for the energy density of the blackbody radiation. In such a
scenario (\ref{densityPhoton}) modifies to,

\begin{align}\label{modifiedModes}
 a(\omega) d\omega &= \frac{1}{\l\pi\r^2} \displaystyle{\sum_{n=0,n^{\prime}=0
\atop n^{\prime}\neq 3}^{\infty}} \frac{a_{n,n^{\prime}}}{n!n^{\prime}!
\kappa^{n+3}} \frac{4\pi}{\left(3-n^{\prime}\right)}
\left[\left(\frac{3V\kappa^3}{4\pi} \right)^{\frac{3-n^{\prime}}{3}}-1 \right]
\omega^{n+2} d\omega \nonumber \\
&+ \frac{1}{\l\pi\r^2}\displaystyle{\sum_{n=0}^{\infty}} \frac{a_{n,3}}{n!3!
\kappa^{n+3}} \bigg(\frac{4\pi}{3}\bigg) \ln\bigg(\frac{3 \kappa^3
V}{4\pi}\bigg) \omega^{n+2} d\omega 
\end{align}
Now the Planck energy density distribution (\ref{distribution}) changes to,

\begin{align}\label{modifiedDistribution} 
 u(\omega) d\omega &= \frac{1}{\l\pi\r^2} \displaystyle{\sum_{n=0,n^{\prime}=0
\atop n^{\prime}\neq 3}^{\infty}} \frac{a_{n,n^{\prime}}}{n!n^{\prime}!
\kappa^{n}} \frac{4\pi}{\left(3-n^{\prime}\right)}
\bigg(\frac{1}{V_{ac}\kappa^3}\bigg) \left[\left(\frac{3V\kappa^3}{4\pi}
\right)^{\frac{3-n^{\prime}}{3}}-1 \right] 
\frac{\omega^{n+3}d\omega}{e^{\frac{\omega}{T}}-1} \nonumber \\
    &+ \frac{1}{\l\pi\r^2}\displaystyle{\sum_{n=0}^{\infty}} \frac{a_{n,3}}{n!3!
\kappa^{n}} \bigg(\frac{4\pi}{3\kappa^3 V_{ac}}\bigg) \ln\bigg(\frac{3 \kappa^3
V}{4\pi}\bigg) \frac{\omega^{n+3}d\omega}{e^{\frac{\omega}{T}}-1} 
\end{align}
\begin{figure}
\centering
\begin{subfigure}{.5\textwidth}
  \centering
  \includegraphics[width=1\linewidth]{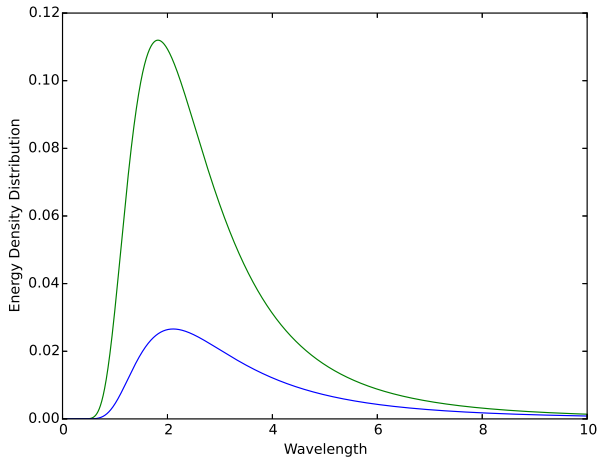}
  \caption{Modified Planck's Distribution}
  \label{fig:pc}
\end{subfigure}%
\begin{subfigure}{.5\textwidth}
  \centering
  \includegraphics[width=1\linewidth]{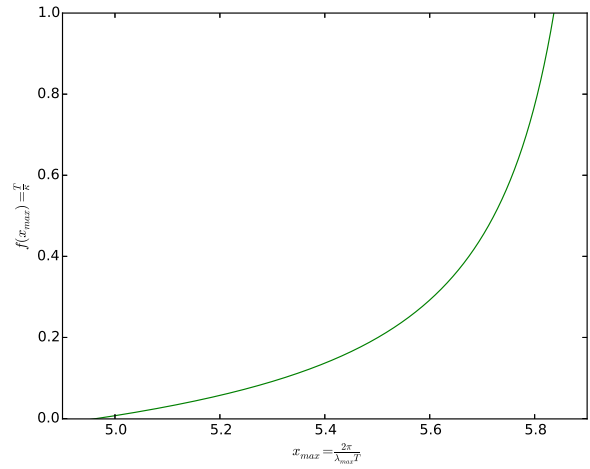}
  \caption{Plot Showing The Modified Nature Of Wien's Law}
  \label{fig:wiens_law}
\end{subfigure}

\caption{\footnotesize{The plot (a) shows the Planck energy distribution as a
function of wavelength $\lambda$ for both modified and unmodified measures. Here
the blue and the green colour correspond to the cases of unmodified and the
modified measures respectively. We have taken $a_{0,0}= 1.0,$ $a_{0,1}=
a_{1,0}=0.2$ and all other $a'$s are zero, temperature is $T=0.6$, volume is
$10^{35}$ and $\kappa=1$ in Planck units. The plot (b) shows the modification in
the Wien's displacement law. The usual Wien's law would have given us a constant
value of $x_{max}$ corresponding to the point where $f(x_{max})=0$. We can
clearly see from the plot that this value is $x_{max} \approx 4.965$. The
coefficient $a_{1,0}$ has been taken to be 1. We have chosen the range such that
$\frac{T}{\kappa}$ lies between $0$ and $1$. Although to decide the actual range
of the possible T-values one needs more thorough and careful thinking, on the
other hand in this paper for our purpose we have discussed the only cases where
T lies 
between $0$ and $\kappa=1$.}}
\label{fig:modified}
\end{figure}
A typical plot of the modified energy density distribution in comparison to the
usual Planck distribution is shown in figure~\ref{fig:pc} on
page~\pageref{fig:pc}.
Let us first express the above distribution in terms of wavelength $\lambda$. 
The energy density between $\omega$ and $\omega+d\omega$ or the corresponding
$\lambda$ and $\lambda+d\lambda$ is $u(\lambda) d\lambda = u(\omega) d\omega$
which implies $u(\lambda) = u(\omega) \frac{d\omega}{d\lambda}=- \frac{\omega^2
u(\omega)}{2\pi}$, 
where $\omega$ and $\lambda$ are related by $\omega = \frac{2\pi}{\lambda}$. 
From (\ref{modifiedDistribution}) we can write $u(\omega) = \sum_{n=0}^{\infty}
A_n \,\, \frac{\omega^{n+3}}{e^{\frac{\omega}{T}}-1}$
where,

\begin{align}
A_n &= \frac{1}{\pi^2} \sum_{n'=0,n'\neq 3}^{\infty} \frac{a_{n,n'}}{n! n'!
\kappa^{n}} \left(\frac{4\pi}{3-n'}\right)\frac{1}{\kappa^3
V_{ac}}\left[\left(\frac{3V\kappa^3}{4\pi} \right)^{\frac{3-n^{\prime}}{3}}-1
\right] \nonumber \\
 &+ \frac{1}{\pi^2} \frac{a_{n,3}}{n! 3! \kappa^{n}} \left(\frac{4\pi}{3\kappa^3
V_{ac}}\right) \ln \left(\frac{3V\kappa^3}{4\pi}\right) 
\end{align}
is a constant and is independent of both $\lambda$ and $T$. We then have,

\begin{equation}
u(\lambda) = -\sum_{n=0}^{\infty} \frac{(2\pi)^{n+4} A_n}{\lambda^{n+5}
\left(e^{\frac{2\pi}{\lambda T}} - 1\right)}.
\end{equation}
Differentiating with respect to $\lambda$ we get

\begin{equation}
\frac{du(\lambda)}{d\lambda} = \frac{1}{\lambda^6\left(e^{\frac{2\pi}{\lambda
T}}-1\right)} \sum_{n=0}^{\infty} \frac{(2\pi)^{n+4} A_n}{\lambda^n}
\left[n+5-\frac{\left(\frac{2\pi}{\lambda T}\right)}{1-e^{-\frac{2\pi}{\lambda
T}}}\right].
\end{equation}
Note that in the case of unmodified measure, we have $A_0 = 1/\pi^2,
A_1=A_2=...=0$
and the above expression reduces to

\begin{equation}
\frac{du(\lambda)}{d\lambda} = \frac{1}{\lambda^6\left(e^{\frac{2\pi}{\lambda
T}}-1\right)} \frac{(2\pi)^{4}}{\pi^2} \left[5-\frac{\left(\frac{2\pi}{\lambda
T}\right)}{1-e^{-\frac{2\pi}{\lambda T}}}\right].
\end{equation}
Thus $u(\lambda)$ is maximum at $\lambda=\lambda_{max}$ which can be found by
the extremum condition $\frac{du(\lambda)}{d\lambda} \bigg|_{\lambda_{max}}=0$
giving $5 = \frac{x_{max}}{1-e^{-x_{max}}}$
where $x_{max}=\frac{2\pi}{\lambda_{max} T}$. The above equation can be
numerically solved to get $x_{max} = \frac{2\pi}{\lambda_{max} T} \approx 4.965,
\,\, \Rightarrow \lambda_{max} T \approx 1.266 $.
This behaviour of $\lambda_{max}$ on temperature $T$ is called {\it Wien's
displacement law}. Now for the case of modified measure the extremum condition
becomes,

\begin{equation}
\sum_{n=0}^{\infty} T^n x_{max}^n A_n
\left[n+5-\frac{x_{max}}{1-e^{-x_{max}}}\right] = 0.
\end{equation}
It is obvious that the solution of $x_{max}$ is now dependent on $T$. Thus the
value of $x_{max}= \frac{2\pi}{\lambda_{max} T}$ is no more constant, but a
function of $T$.
To understand the behaviour in a better way, we keep the leading order terms in
$\frac{T}{\kappa}$ and $\frac{1}{V^{1/3}\kappa}$ and neglect all the higher
order terms i.e. $A_0 \approx 1/\pi^2, \quad A_1 \approx \frac{a_{1,0}}{\pi^2
\kappa}, \quad A_2 \approx A_3 \approx ... \approx 0$.
The extremum condition then becomes

\begin{equation}
\frac{T}{\kappa} = -\frac{1}{ x_{max} a_{1,0} }
\left(\frac{5-\frac{x_{max}}{1-e^{-x_{max}}}}{6-\frac{x_{max}}{1-e^{-x_{max}}}}
\right) = f(x_{max}).
\end{equation}
We have plotted this function with respect to $x_{max}$ (see
figure~\ref{fig:wiens_law} on page~\pageref{fig:wiens_law}). For a fixed value
of $y-$axis, i.e., a fixed $\frac{T}{\kappa}-$ value the corresponding value of
$x_{max}$ can be obtained from the plot.
As visible from the plot $x_{max}=\frac{2\pi}{\lambda_{max}T}$ is a
monotonically increasing function of $T$, i.e., $f^{-1}(\frac{T}{\kappa})$. This
implies $\lambda_{max}=\frac{2\pi}{Tf^{-1}(\frac{T}{\kappa})}$ is a
monotonically decreasing function of $T$. Note that $\lambda_{max}$ for modified
phase space measure decreases
more rapidly with increasing $T$ than the case of unmodified measure where
$x_{max}=f^{-1}(\frac{T}{\kappa})$ takes a constant value.
The significant change in the values of $x_{max}$ occurs only if the order of
the change in temperature is non-negligible with respect to $\kappa$. That is
why in SR limit, i.e., $\frac{T}{\kappa} \rightarrow 0$, the $x_{max} =
\frac{2\pi}{\lambda_{max} T}$ is almost constant giving the standard Wien's
displacement law. The 
extremum condition for the unmodified measure corresponds to $f(x_{max})=0$. As
it is visible in figure~\ref{fig:wiens_law} on page~\pageref{fig:wiens_law} this
gives the usual value $x_{max} \approx 4.965$. 
Note that the value of $f^{-1}\left(\frac{T}{\kappa}\right)$ is always greater
than the SR value $4.965$. Hence
$\frac{(\lambda_{max})_{DSR}}{(\lambda_{max})_{SR}}=\frac{4.965}{f^{-1}
\left(\frac{T}{\kappa}\right)} \leq 1$.
Thus, the frequency at which the energy density distribution
of blackbody radiation at a given temperature peaks, gets a positive correction.
Now, suppose we demand at least $1\%$ correction i.e.,
$\frac{(\lambda_{max})_{DSR}}{(\lambda_{max})_{SR}}=\frac{4.965}{f^{-1}
\left(\frac{T}{\kappa}\right)}=\frac{99}{100}$ then we get
$f^{-1}\left(\frac{T}{\kappa}\right)= 5.015$. The corresponding
$\frac{T}{\kappa}$ from the plot is $0.01$.
So, to get an observable effect of DSR using modified Wien's displacement law
one needs to consider a system having temperature in the range of $100$th part
of the effective $\kappa$ value. 
Note that some exotic phenomenon in the semi-classical regime of quantum gravity
may reduce the effective value of the energy cut-off in certain specific
systems. A similar reduction in the effective value of high energy cut-off has
been suggested in a simple quantum mechanical table top experiment in section
\ref{physical}.
\begin{figure}
\centering
\begin{subfigure}{.4\textwidth}
  \centering
  \includegraphics[width=.90\linewidth]{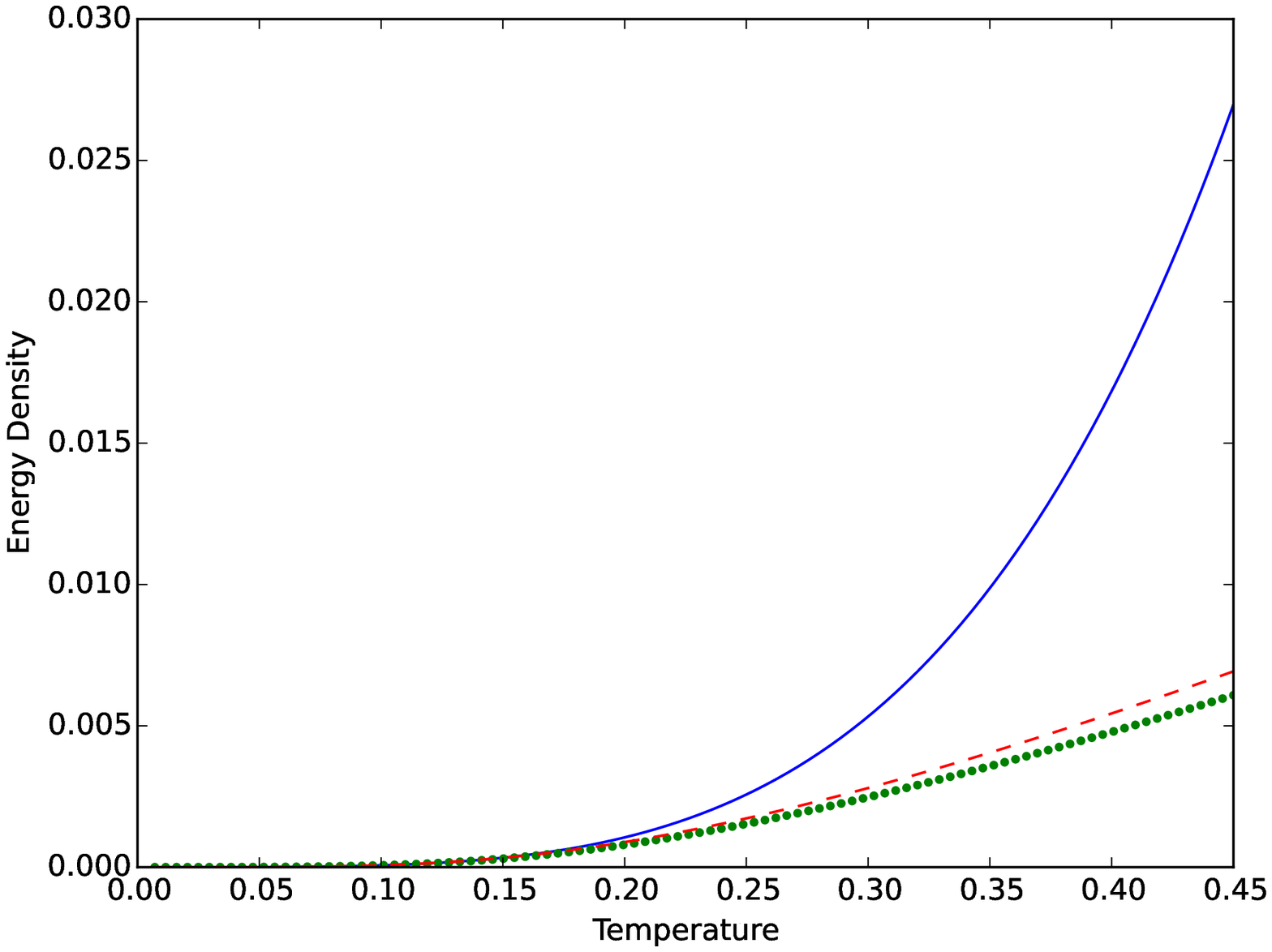}
  \caption{Energy Density Distribution}
  \label{fig:ed}
\end{subfigure}%
\begin{subfigure}{.4\textwidth}
  \centering
  \includegraphics[width=.90\linewidth]{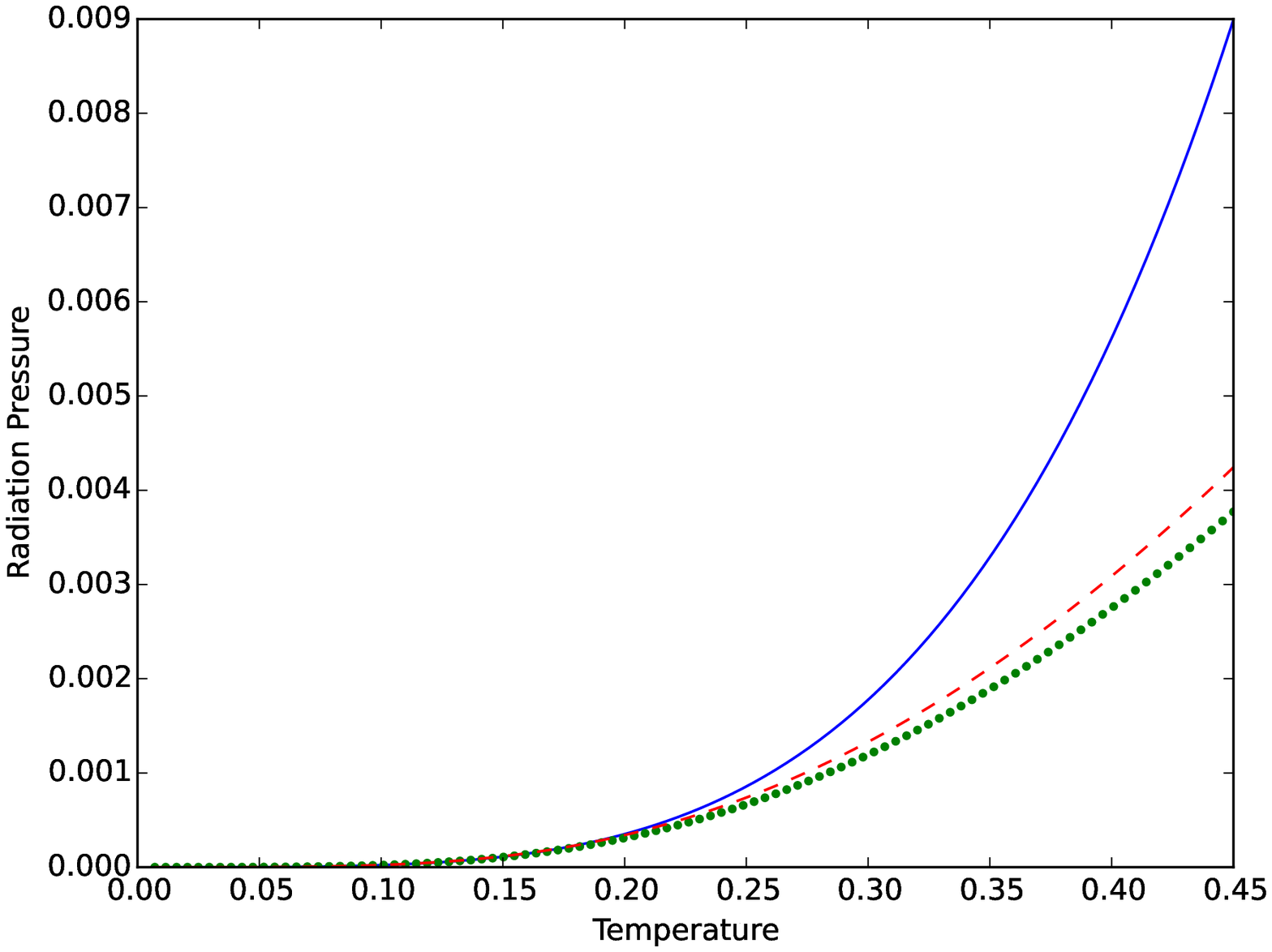}
  \caption{Radiation Pressure}
  \label{fig:rp}
\end{subfigure}
\begin{subfigure}{.4\textwidth}
  \centering
  \includegraphics[width=.90\linewidth]{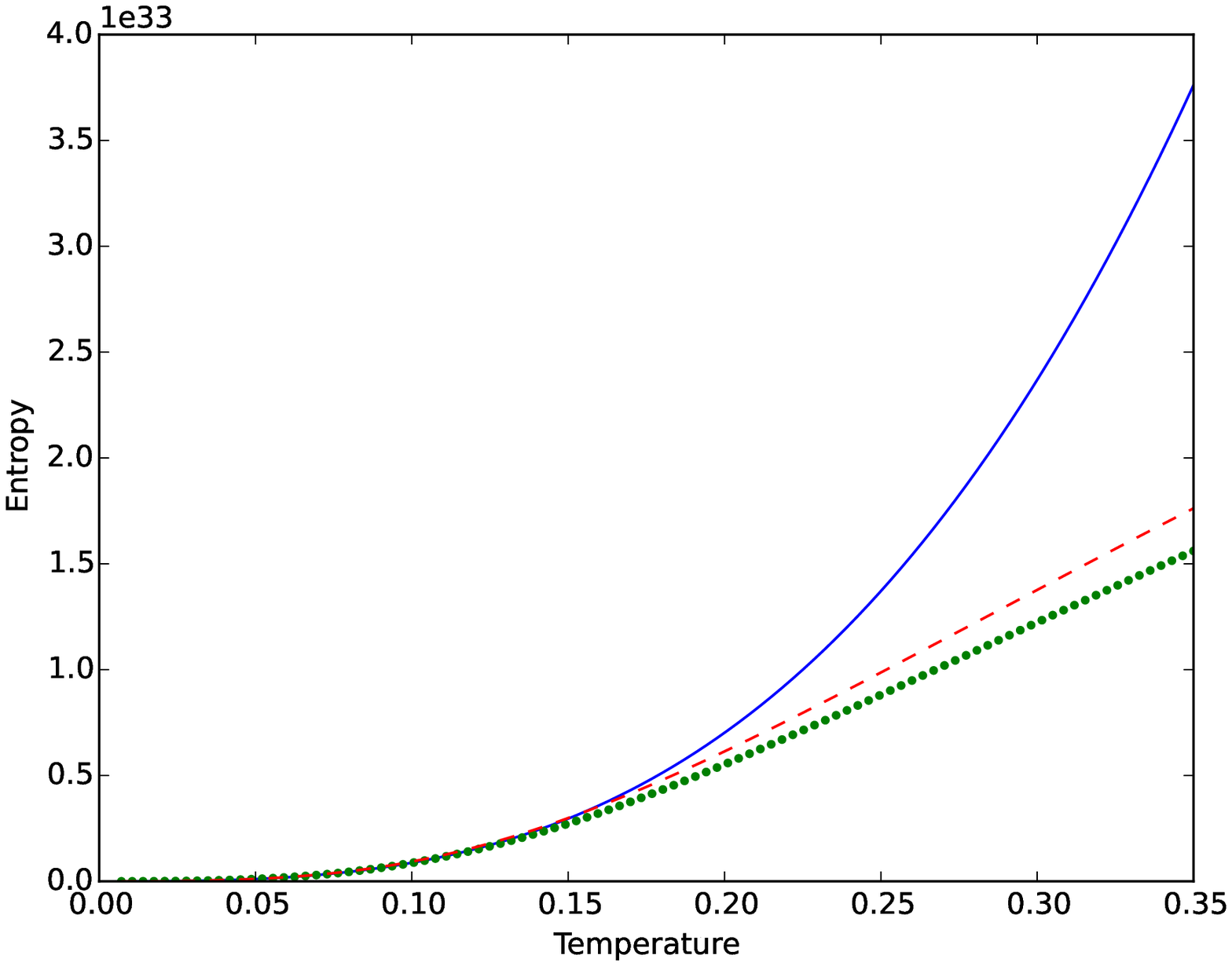}
  \caption{Entropy}
  \label{fig:ent}
\end{subfigure}%
\begin{subfigure}{.4\textwidth}
  \centering
  \includegraphics[width=.90\linewidth]{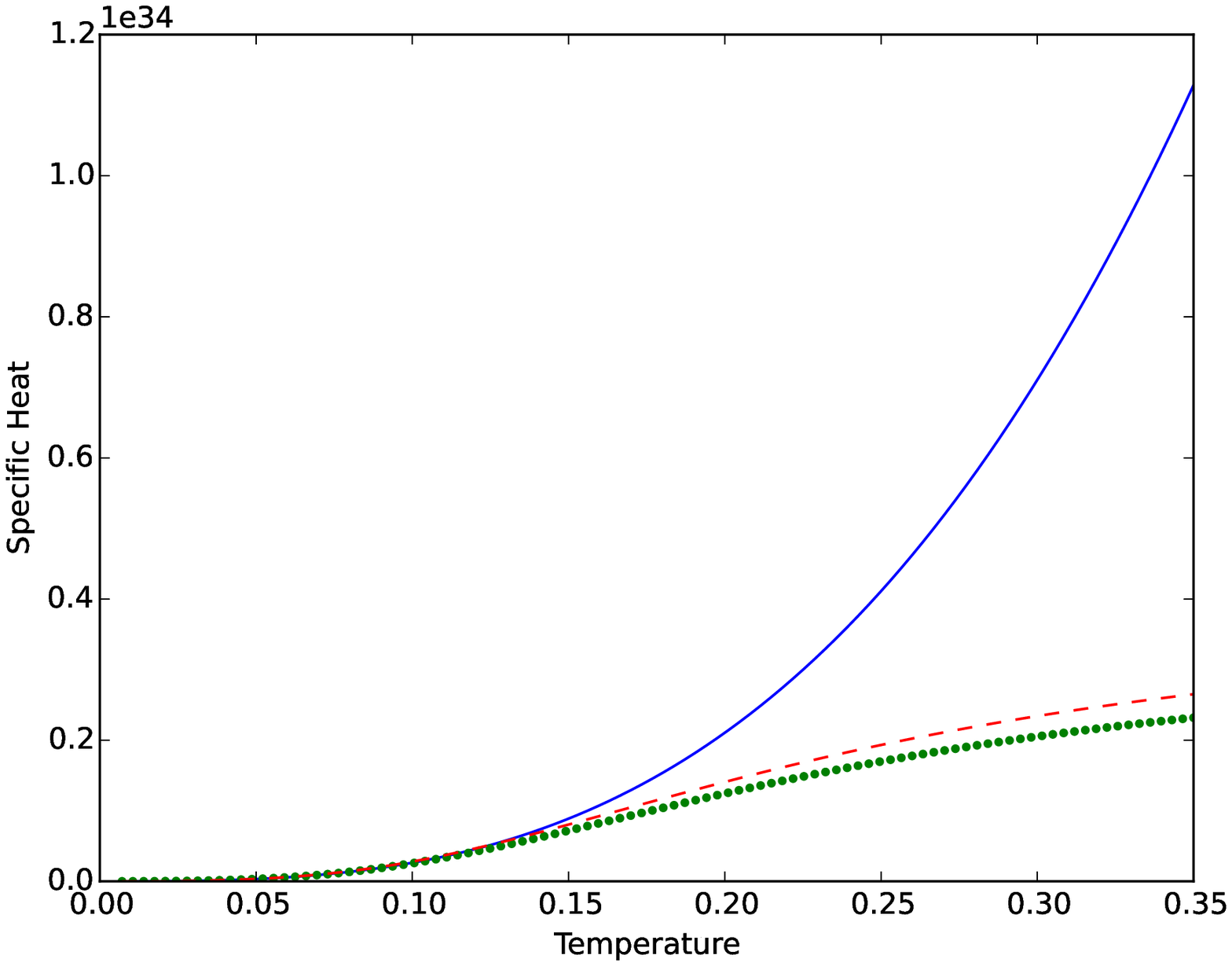}
  \caption{Specific Heat}
  \label{fig:sh}
\end{subfigure}
\begin{subfigure}{.4\textwidth}
  \centering
  \includegraphics[width=.90\linewidth]{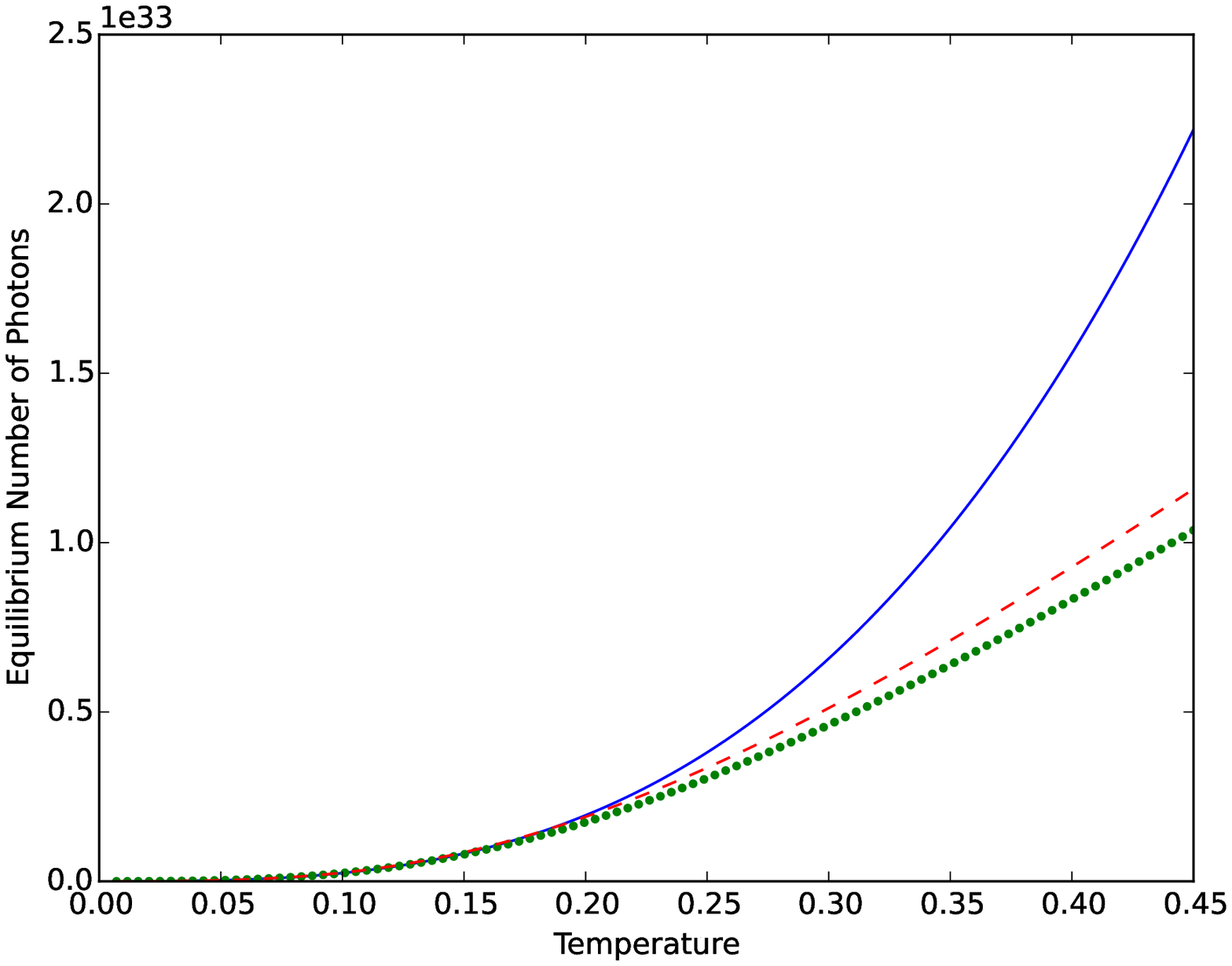}
  \caption{Equilibrium Number Of Photons}
  \label{fig:pn}
\end{subfigure}%
\begin{subfigure}{.4\textwidth}
  \centering
  \includegraphics[width=.90\linewidth]{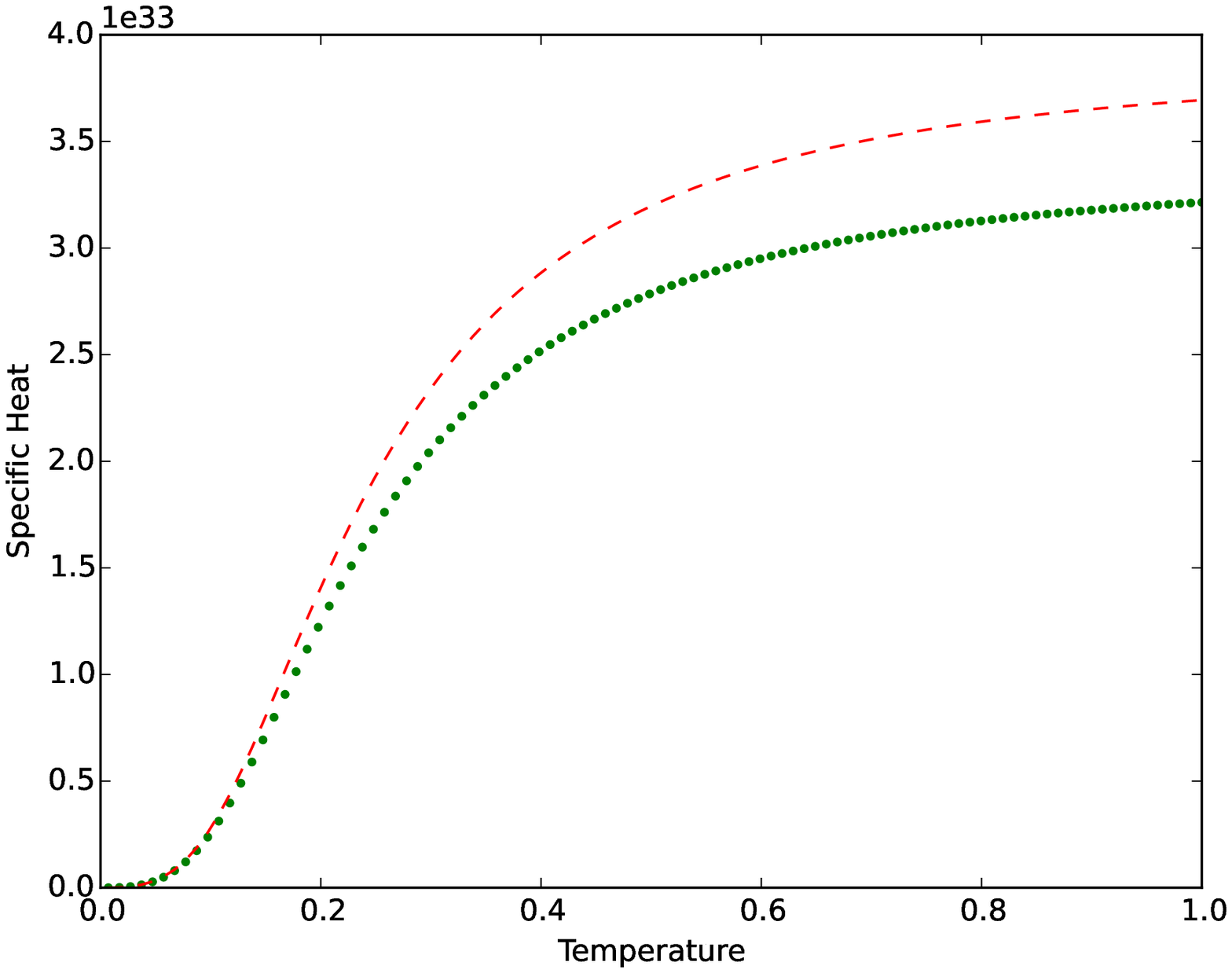}
  \caption{Specific Heat Mimicking Debye Theory}
  \label{fig:sh1}
\end{subfigure}%
\caption{\footnotesize{The plots show the variation of energy density, radiation
pressure, entropy, specific heat and equilibrium number of photons with
temperature for thermodynamic quantities with an ultraviolet energy cut-off and
also
with the modified measure. 
The blue solid, the green dotted and the red dashed lines correspond to the SR,
the case with the ultraviolet energy cut-off and the case with the modified
measure respectively. 
As is visible from the plots it matches with SR at low T and with increasing T
it deviates from SR significantly. Here $\kappa=1$ and $V=10^{35}$ in Planck
units and $a_{0,0}=1$,
$a_{0,1}=a_{1,0}=0.2$ and all other $a'$s are taken to be zero. 
Note that  all the quantities above become approximately linear near $T
\rightarrow \kappa$. In the low temperature regime, energy density $u$ and
radiation pressure $P$ follow $\sim T^4$ behaviour, while the entropy $S$, the
specific heat $C_V$ and the equilibrium number of photons $\bar{N}$ follow $\sim
T^3$ behaviour. 
The behaviour of $C_V$ for the full range of $\frac{T}{\kappa} \in [0,1]$ is
shown in the figure at bottom right corner, which certainly mimics the Debye
theory. 
In the Debye theory however T may go up to infinity in which case the specific
heat goes to a constant value.} }
\label{fig:plot}
\end{figure}
\subsubsection{Various thermodynamic quantities with modified measure}
We have calculated various thermodynamic quantities with the modified measure.
The exact results in form of lengthy expressions are listed in \ref{modified}
and here we proceed further with physical analysis only.
Note that in this case all the thermodynamic quantities reduce to the unmodified
case for
$n=0$ and $n^{\prime}=0$. Also in the SR limit, we get the usual SR result as
expected. The value of the thermodynamic quantities,
in this case, can either be less than or equal to (for certain T-values only) or
greater than both the SR value and the values in case with only an ultraviolet
energy cut-off,
depending on the choice of $a_{n,n'}$. To plot these quantities we have chosen
the $a_{n,n'}$ values in such a way that
the value of the modified case is more than the unmodified case and less than
the SR case. Though it is not visible in the plot because of the chosen
$a_{n,n'}$ values, but it is a fact that for certain choices of $a_{n,n'}$ the
modified DSR value becomes equal to the value of SR at some temperatures and can
even overshoot the SR curve.
The nonanalytic nature in the SR limit for the case of modified measure is
similar as in the unmodified case.
The leading order behaviour in the low and high temperature limits are discussed
in the next section.
\section{Effects of DSR in Big Bang and cosmology}
In this section we will explore the possible effects of the behaviour of DSR
photons near the Planck scale
with an invariant ultraviolet energy cut-off. 
To see the physical 
applicability of the results with such an invariant ultraviolet energy cut-off
one has to, in general, probe near the Planck
scale. The results can then be used to study the
early Universe thermodynamics especially Big Bang cosmology. Since we have the
modified energy density $u$ and Pressure $P$, therefore we have a modified
energy-momentum tensor $T_{\mu \nu}$.
In general, we should use the modified metric when we are exploring the early
Universe near Planck scale. In DSR, Smolin has suggested one such metric called
the Rainbow metric
\cite{Magueijo:2002xx}\cite{AmelinoCamelia:2003xp}\cite{Assaniousssi:2014ota}
\cite{Loret:2015iia}\cite{ Assanioussi:2016yxx}, but we will not attempt to
discuss this here. 
With the above quantities at hand, we can then solve the
Friedmann equations (more specifically FRW equations) 
and see the possible modification in the known results of the expansion of the
Universe after Big Bang at such a scale. This is very involved and a more
detailed
study can be done separately in future.
But we can still consider a scenario where we can see the possible modification
near the Big Bang. 
The FRW and its relation is a standard and well studied cosmology subject. We,
for our analysis, will follow chapter 8 of \cite{carroll}. 
We will consider the radiation dominated epoch where the
modified energy density and pressure is given by (\ref{energyDensity}) and
(\ref{PressureUnmod}) respectively. With such a modification of $T_{\mu \nu}$,
the energy
conservation equation (8.54) in section 8.3 of \cite{carroll} gets modified to
\begin{align}
 \frac{\dot{u}}{u}=-\left( 4- \frac{\frac{T{\kappa}^3}{{\pi}^2}
ln(1-e^{-\frac{\kappa}{T}})}{u} \right) \frac{\dot{a}}{a}.
\end{align}
We then express $u$ in terms of $T$ to get,
\begin{align}
H=\frac{\dot{a}}{a}=-\frac{\dot{T}}{T}\left[
\frac{24\left[Z_4(0)-Z_4\left(\frac{\kappa}{T}\right)\right]-\left(\frac{\kappa}
{T}\right)^4
Li_0\left(e^{\frac{-\kappa}{T}}\right)}{24\left[Z_4(0)-Z_4\left(\frac{\kappa}{T}
\right)\right]+\left(\frac{\kappa}{T}\right)^3
Li_1\left(e^{\frac{-\kappa}{T}}\right)}\right] 
\end{align}
Here $a$ is the dimensionless scale factor and $H$ is the Hubble parameter which
characterizes the rate of expansion of
the Universe.
It is easy to see that the numerator is always less than the denominator. 
Therefore $\frac{H}{H_{SR}}<1$ always, where $H_{SR}=-\frac{\dot{T}}{T}$, which
implies
that the expansion of the Universe was at a slower rate in the radiation
dominated
era than the rate of expansion without such modifications. Because of the slower
expansion, all the epochs would eventually get delayed resulting in the
modification in the age of the known Universe.  
In the SR limit the modified Hubble parameter becomes nearly
equal to the normal SR one,
as $\frac{\kappa}{T}$ is very large, so the correction terms go to zero as
expected. 

We can also see its application in case of ``bouncing" loop quantum cosmology
theories (see \cite{Ashtekar:2008ay}\cite{Bojowald:2010xq}\cite{Alesci:2016xqa}
and the references therein), where normally we consider specific modifications
to the spacetime
geometry which effectively puts a bound on the curvature and in this way the Big
Bang singularity can be avoided. 
But for such “bouncing” models, we cannot use the perturbation technique at the
curvature saturation, as the energy density of the cosmic fluid diverges. 
What one can do to still avoid the Big Bang singularity is to consider an
inflation
model where we can safely use the perturbation theory. 
Here we have obtained the energy density of the cosmic fluid which saturates to
the Planck energy which of course is finite. 
Then we can combine both the results obtained in this paper and the ``bouncing"
loop quantum cosmology to study the possible way out to avoid the Big Bang
singularity.

Next we consider the DSR photons at an effective lower scale due to other
parameters in the theory like mass, number density etc. These parameters may
effectively lower the Planck scale such that its effects can be observed in very
high temperature and high density regimes.
To probe such DSR effects we need to observe the stellar
objects with very high temperatures and densities. 
For example, the astronomical data from
gamma ray burst during the merging of neutron stars (which has the core
temperature of $T=10^{12} $ K) may give a bound on effective $\kappa$ value.
We can also explore the Chandrasekhar limits and its possible modifications.
The application of the theory developed here has been explored in detail for
white dwarfs in \cite{WhiteD}.
It will also be interesting to see if one gets a better bound on $\kappa$ in
case of
luminosity calculation of neutron stars using the results obtained for the
blackbody radiation in this paper. 

\section{The leading behaviour for $T \rightarrow 0$ and $T \rightarrow \kappa
$}\label{leading}
We have plotted various thermodynamic quantities as a function of temperature
(see  figure~\ref{fig:plot} on page~\pageref{fig:plot}).
Let us now analyse the behaviour near $T=0$ and $T=\kappa$. In the low
temperature regime we take $\frac{T}{\kappa}=\epsilon<<1$.
The low temperature behaviour is as follows,

\begin{align}
 u \approx \frac{{\pi}^2 \kappa^4 {\epsilon}^4}{15}- \frac{\kappa^4}{\pi^2}
\epsilon e^{-\frac{1}{\epsilon}} \approx \frac{{\pi}^2 T^4}{15}, \quad P \approx
\frac{{\pi}^2 T^4}{45}, \nonumber \\
 S \approx \frac{4 V_{ac} {\pi}^2 T^3}{45}, \quad C_V \approx \frac{4 V_{ac}
{\pi}^2 T^3}{15}, \quad \bar{N} \approx \frac{2 V_{ac} \zeta(3) T^3}{{\pi}^2}.
\end{align}
Here we have used the fact that $Li_n(z)\rightarrow z$ as $z\rightarrow 0$.
In the expression for energy density, we have neglected the second term with
respect to the first. We can see this by putting $x=\frac{1}{\epsilon}$ and as
$x \rightarrow \infty$ the ratio of the second term to the first, in the above
equation goes to zero.
Note that the second term $\epsilon e^{-\frac{1}{\epsilon}}$ is the nonanalytic
piece which makes this limit non-perturbative i.e. this expression cannot be
Taylor series expanded in the low temperature limit. Let us consider the energy
density
relation in the $\frac{T}{\kappa}=\epsilon<<1$ limit, given by  $u \approx
\frac{{\pi}^2 \kappa^4 {\epsilon}^4}{15}- \frac{\kappa^4}{\pi^2}
\epsilon e^{-\frac{1}{\epsilon}}$. Now assuming that we get at least $1\%$
correction i.e.,

\begin{align}
\frac{\frac{\kappa^4}{\pi^2} \epsilon e^{-\frac{1}{\epsilon}}} {\frac{{\pi}^2
\kappa^4 {\epsilon}^4}{15}} \geq \frac{1}{100} 
\end{align}
which, in turn, gives a bound on $\epsilon$ as $0.10\leq \epsilon \leq
2.1$.
But since we have taken $\epsilon<<1$, therefore the equality holds at
$\epsilon=0.10$. This fact is also visible from the plots of the thermodynamic
quantities in which the modified behaviour starts deviating from the SR result 
at
$\frac{T}{\kappa} \sim 10^{-1}$.
Another point to note is that the value of $\epsilon$, for at least $1\%$
correction, in case
of modified Wien's displacement law came out to be around $0.01$ (See section
\ref{modWien}). Thus, the modified Wien's displacement law 
starts giving an observable correction for the systems having temperature one
order less compared to the systems used in case of modified thermodynamic
quantities.
For pressure we do the similar analysis where the first term in
(\ref{PressureUnmod}) is nothing but $\left(\frac{\kappa^3 T}{3 \pi^2}\right)
Li_1(e^{\frac{-\kappa}{T}})$. A similar analysis follows for other thermodynamic
quantities as well.
Thus in low temperature regime, energy density $u$ and radiation pressure $P$
follow $\sim T^4$ behaviour, while the entropy $S$, the specific heat $C_V$ and
the equilibrium number of photons $\bar{N}$ follow $\sim T^3$ behaviour. The
nonanalyticity
in this limit is a general feature of all the thermodynamic quantities.
For high temperature, $T\approx\kappa(1-\epsilon)$  such that $\epsilon<<1$
which gives $\frac{\kappa}{T}\approx1+\epsilon$. We will expand all the
quantities to the leading order in $\epsilon$ and finally put $\epsilon
=1-\frac{T}{\kappa}$ to get the leading high temperature behaviour. The results
are listed in \ref{unmodified_limit}.
Note that to get the linear dependence of $C_V$ on $T$ in (\ref{C_V_limit}) by
differentiating the high $T$ behaviour of $U$, we need to expand $U$ up to $T^2$
order.
All these linear behaviours for $T \rightarrow \kappa$ are very clearly visible
in the plots.
For the modified measure we get essentially the similar behaviour for both the
limits. The results of the leading behaviour in case of modified measure are
listed in \ref{modified_low_limit} and \ref{modified_high_limit}.
\begin{figure}
\centering
  \includegraphics[width=0.4\textwidth]{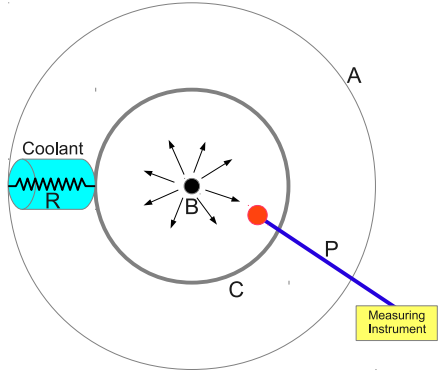}
\caption{\footnotesize{A quantum mechanical table top experiment for testing the
theory. Here B is a perfect blackbody surrounded by a spherical
cathode C, which is enclosed by a spherical
anode A and the circuit is completed using a high resistance resistor R.
All the radiations above the frequency $\nu_{th}$ corresponding to the
photoelectric
threshold of the cathode are absorbed and will be lost as Joule heating through
R. The resistor is continuously being cooled by coolant, shown in blue colour, to
avoid the melting of the resistor. The
photons below the frequency $\nu_{th}$ stay inside the cavity surrounded by the
cathode C and behave as acoustic
phonons as
shown in this paper. The properties of these photons acting as acoustic phonons
can be
measured by inserting a probe P connected with the measuring instrument.}}
\label{fig:tableTop}
\end{figure}

\section{Effective low energy realizations of the theory}\label{physical}

In this section, we present the possibilities of physical
realization of the results obtained with an effective cut-off for the photons
such that they behave as phonons. As is clear
from the description this cut-off might not be invariant which was the case in
the other applications discussed above. But this is an interesting case in its
own regard, as we have a way to get a bunch of photons behaving as phonons and
they can be observed in a laboratory as well.
To start with here we will be suggesting a simple table top experiment to test
the result
obtained in section \ref{unmodified}. Note that due to no change in dispersion
relation the
only effect on the thermodynamics of a photon gas is due to the high energy
cut-off. If one can introduce such a cut-off on the photon energy in some
experiment then the photons will start behaving like acoustic phonons.
Consider a perfect blackbody (see figure~\ref{fig:tableTop} on
page~\pageref{fig:tableTop}) surrounded by a spherical cathode which is
enclosed by a spherical anode and the 
circuit being completed using a high resistance. Now, suppose
the cathode has the photoelectric threshold $\nu_{th}$ such that all the
radiations above
frequency $\nu_{th}$ gets absorbed by
the cathode. These absorbed radiations lead to the Joule heating of the resistor
which is then cooled by an appropriate coolant. 
Since, we do not want the heating of the resistor, in any way, to affect the
radiations inside the cathode, therefore the cathode may be coated
with an insulating material.
Another possible alternative
is to drill a small hole in anode then connect the resistor outside and far
away from the anode where it can be cooled. Now, we are left with photons in the
cathode cavity, with energy less than $\kappa=\nu_{th}$ which is the desired
cut-off in
the theory. 
We, therefore, have generated
photons inside the cathode which mimic acoustic phonons. To observe this we can
now drill a
very small hole in both anode and cathode, through which a probe,
connected to the measuring instrument, can be inserted to test the properties of
the photons. 
To observe at least $1\%$ deviation (see the discussion in section
\ref{leading})
from the usual photon thermodynamics at room temperature $T=293$ K the material
of the cathode can be selected
with threshold $\nu_{th}=\frac{k_B T}{0.1 h} = 6.1 \times 10^{13}
Hz$ (here we have put the actual
values of $h$ and $k_B$). Many commercially available
materials fall into this category.

\section{Summary and future works}
We started with the DSR formalism by MS where the modified dispersion relation,
in order to incorporate an invariant energy scale $\kappa$, is given by
(\ref{MS}). But in the case of photon gas, it is simply $\e=p$.
Since there is a cut-off on the maximum energy and the minimum length, the
expression of the thermodynamic quantities changes accordingly. 
We started by considering a model of a photon gas obeying Bose-Einstein
statistics in grand canonical ensemble and went on calculating various
thermodynamic quantities such as energy density, pressure, entropy, specific
heat and equilibrium number of photons with such an ultraviolet cut-off. We
found one to one correspondence between
the behaviour of photons with an ultraviolet cut-off and the acoustic phonons in
the Debye theory. The
Stefan-Boltzmann law got modified which will give correction to the dynamics of
many stellar objects. We found that the non-perturbative nature of the
thermodynamic quantities in 
the SR limit is a general feature of the theory with an ultraviolet energy
cut-off. We also noted that the values of all the thermodynamic quantities are
less than the SR values because of this cut-off.
We then studied the change in the phase space measure for exotic spacetimes at
Planck scale and discussed the example of classical ideal gas for illustration.
We found that the classical ideal gas in case of modified phase space measure
has a non-trivial volume dependence in its expression for the partition function
leading to the modification in the thermodynamic quantities like pressure
accordingly.
We went on calculating the possible change in the thermodynamic quantities due
to the change in the phase space measure. Because of this modification, Planck's
energy density distribution and the Wien's displacement law got modified.
Note that all the thermodynamic quantities reduce to the usual SR result in
$\kappa \rightarrow \infty$ limit. We have plotted the temperature dependence of
various thermodynamic quantities. We can clearly see from the plots of various
thermodynamic quantities that we
start getting the deviation of the results obtained in the modified case from
the SR case at
$\frac{T}{\kappa} \sim 10^{-1}$. We found that the
modified Wien's law can be observed at comparatively lower temperature than
the thermodynamic quantities.
Next, we discussed the possible realization of the modification at Planck
scale by considering its effects near the Big Bang.
The effectively lower Planck scale cosmological observations and 
modifications of DSR have also been discussed.
The leading behaviour for $T \rightarrow 0$ and $T \rightarrow \kappa$ have been
analysed. We observed that in the case of modified phase space measure the
values of the thermodynamic
quantities might be less than, equal to or greater than the SR values
depending on the choice of $a_{n,n'}$.
As was seen both for the case with only an ultraviolet cut-off and the modified
measures, the
nonanalyticity in the special relativistic limit is a general feature of the
energy cut-off introduced in the theory.
In the last section, we have given the possible scenarios of the
physical observation of results obtained in effective low energy by
suggesting a quantum mechanical table top experiment.
Note that the present work only deals with the study of massless bosons i.e.
photons, but the analysis can be extended to massive bosons and fermions as
well. This will give us an insight into the invariant energy scale effects in
well-known phenomenon such as Bose-Einstein condensation and the behaviour of
degenerate Fermi gas.  
This, in turn, leads to the study of the stellar objects such as white dwarfs
and
neutron stars. It would certainly be interesting to see whether the
non-perturbative effects appear in such massive cases too. Analysis of this
paper can be used to study early Universe
thermodynamics.

\section*{Acknowledgements}
Authors would like to thank the referee for useful comments and suggestions.
DKM would like to thank Prashanth Raman, Sanjoy Mandal, Anirban Karan, Pritam
Sen and others for various useful discussions.

\begin{appendices}
\section{Criterion for swapping the double summation and the integral}
\label{App}
The interchange of summation and integration signs for a series is a well known
theorem (see theorem 1.38 of \cite{rudin}). In this Appendix, we will generalize
this theorem for a double series i.e. we will prove the equality,

\begin{align}\label{equality}
 \int_{\mathcal{M}} \sum_n \sum_{n^{\prime}} A_{n,n^{\prime}}=\sum_n
\sum_{n^{\prime}}\int_{\mathcal{M}} A_{n,n^{\prime}}
\end{align}
holds if, 

\begin{align}\label{condition1}
 \sum_n \sum_{n^{\prime}}\int_{\mathcal{M}}  |A_{n,n^{\prime}}| < \infty.
\end{align}
We take the two summations out of the integral one by one in the LHS of
(\ref{equality}) to get the RHS, using theorem 1.38 of \cite{rudin} which is
allowed if,

\begin{align}\label{condition}
 \sum_n \int_{\mathcal{M}}  \left|\sum_{n^{\prime}} A_{n,n^{\prime}}\right| <
\infty \quad \rm{and} \quad
\sum_{n^{\prime}}\int_{\mathcal{M}} | A_{n,n^{\prime}}|  < \infty \quad \forall
n.
\end{align}
Let us rewrite (\ref{condition1}) as,

\begin{align}\label{tn}
  \sum_n t_n < \infty \quad \rm{with} \quad t_n=\sum_{n^{\prime}}
\int_{\mathcal{M}} |A_{n,n^{\prime}}| \quad \forall n.
\end{align}
We note that $t_n\geq0 \quad \forall n$, which along with (\ref{tn}) implies
$t_n< \infty \quad \forall n $ which is nothing but the second inequality in
(\ref{condition}). This further implies (because of theorem 1.38 of
\cite{rudin}),

\begin{align}\label{tnNew}
 t_n=\int_{\mathcal{M}} \sum_{n^{\prime}} |A_{n,n^{\prime}}| \quad \forall n.
\end{align}
Now we note that $
 \left|\sum_{n^{\prime}} A_{n,n^{\prime}} \right|\leq \sum_{n^{\prime}} 
|A_{n,n^{\prime}} | \quad \forall n$.
Integrating over $\mathcal{M}$ followed by the summation over $n$ we get,

\begin{align}
\sum_{n} \int_{\mathcal{M}}  \left|\sum_{n^{\prime}} A_{n,n^{\prime}} \right|
&\leq \sum_{n} \int_{\mathcal{M}} \sum_{n^{\prime}} | A_{n,n^{\prime}}| \quad
\forall n
\end{align}
Taking (\ref{tnNew}) and (\ref{tn}) into account in the above inequality we get
the first inequality of (\ref{condition}). Thus we have proved that
(\ref{condition1}) implies (\ref{condition}) and hence if (\ref{condition1}) is
satisfied then the equality (\ref{equality}) holds true.
\section{List of certain results}
\subsection{Thermodynamic quantities with modified measure}\label{modified}
Following what we did in section \ref{energySection} along with using
(\ref{modifiedDistribution}) we get,

\begin{align}\label{modifiedU}
   u &= \frac{1}{\l\pi\r^2} \displaystyle{\sum_{n=0,n^{\prime}=0 \atop
n^{\prime}\neq 3}^{\infty}} \frac{a_{n,n^{\prime}}}{n!n^{\prime}! \kappa^{n}}
\frac{4\pi}{\left(3-n^{\prime}\right)}
\bigg(\frac{T^{n+4}}{V_{ac}\kappa^3}\bigg) \left[\left(\frac{3V\kappa^3}{4\pi}
\right)^{\frac{3-n^{\prime}}{3}}-1 \right] \Gamma(n+4)
\bigg[Z_{n+4}(0)-Z_{n+4}\left(\frac{\kappa}{T}\right)\bigg]  \nonumber \\
   &+ \frac{1}{\l\pi\r^2}\displaystyle{\sum_{n=0}^{\infty}} \frac{a_{n,3}}{n!3!
\kappa^{n}} \bigg(\frac{4\pi T^{n+4}}{3V_{ac}\kappa^3}\bigg) \ln\bigg(\frac{3
\kappa^3 V}{4\pi}\bigg)  \Gamma(n+4)
\bigg[Z_{n+4}(0)-Z_{n+4}\left(\frac{\kappa}{T}\right)\bigg] \nonumber \\
   &=\displaystyle{\sum_{n=0,n^{\prime}=0}^{\infty}} u_{n,n^{\prime}}.
\end{align}
Here $u_{n,n^{\prime}}$ is a general term of the summation. 
Therefore the specific heat capacity $C_V$ is,

\begin{align}\label{modifiedC_V}
 C_V &=\bigg(\frac{\partial U}{\partial
T}\bigg)_{V_{ac}}=\displaystyle{\sum_{n=0,n^{\prime}=0 \atop n^{\prime}\neq
3}^{\infty}} \left [\frac{1}{\l\pi\r^2}  \frac{a_{n,n^{\prime}}}{n!n^{\prime}!
\kappa^{n+3}} \frac{4\pi}{\left(3-n^{\prime}\right)}
\left[\left(\frac{3V\kappa^3}{4\pi} \right)^{\frac{3-n^{\prime}}{3}}-1 \right]
\bigg\{\frac{1}{(1-e^{\frac{\kappa}{T}})} \frac{\kappa^{n+4}}{T} \bigg\}+
(n+4)\left(\frac{u_{n,n^\prime}V_{ac}}{T}\right) \right] \nonumber \\
  &+\displaystyle{\sum_{n=0}^{\infty}} \left[ \frac{1}{\l\pi\r^2}
\frac{a_{n,3}}{n!3! \kappa^{n+3}} \left(\frac{4\pi}{3}\right)
\ln\bigg(\frac{3\kappa^3 V}{4\pi}\bigg)
\bigg\{\frac{1}{(1-e^{\frac{\kappa}{T}})} \frac{\kappa^{n+4}}{T} \bigg\}+
(n+4)\left(\frac{u_{n,3}V_{ac}}{T}\right) \right].
\end{align}
The radiation pressure can easily be calculated in essentially the similar
manner to get,

\begin{align}
  P &=\frac{1}{\l\pi\r^2} \displaystyle{\sum_{n=0,n^{\prime}=0 \atop
n^{\prime}\neq 3}^{\infty}} \frac{a_{n,n^{\prime}}}{n!n^{\prime}! \kappa^{n}}
\frac{4\pi}{\left(3-n^{\prime}\right)} \bigg(\frac{T}{V_{ac}\kappa^3}\bigg)
\left[\left(\frac{3V\kappa^3}{4\pi} \right)^{\frac{3-n^{\prime}}{3}}-1 \right]
\bigg\{ -\ln(1-e^{-\frac{\kappa}{T}}) \frac{\kappa^{n+3}}{n+3}
  + \frac{T^{n+3}}{(n+3)} \Gamma(n+4)
\bigg[Z_{n+4}(0) \nonumber
\\ 
   & -Z_{n+4}\left(\frac{\kappa}{T}\right)\bigg] \bigg\} +\frac{1}{\l\pi\r^2}\displaystyle{\sum_{n=0}^{\infty}} \frac{a_{n,3}}{n!3!
\kappa^{n}} \bigg(\frac{4\pi T}{3V_{ac}\kappa^3}\bigg) \ln\bigg(\frac{3 \kappa^3
V}{4\pi}\bigg) \bigg\{ -\ln(1-e^{-\frac{\kappa}{T}}) \frac{\kappa^{n+3}}{n+3} +
\frac{T^{n+3}}{(n+3)} \Gamma(n+4)
\bigg[Z_{n+4}(0)-Z_{n+4}\left(\frac{\kappa}{T}\right)\bigg] \bigg\}.
 \end{align}
This can be related to the energy density as

\begin{align}\label{modifiedP}
P   &= \displaystyle{\sum_{n=0,n^{\prime}=0 \atop n^{\prime}\neq 3}^{\infty}}
\left[ \frac{1}{\l\pi\r^2} \frac{a_{n,n^{\prime}}}{n!n^{\prime}! \kappa^{n}}
\frac{4\pi}{\left(3-n^{\prime}\right)} \bigg(\frac{T}{V_{ac}\kappa^3}\bigg)
\left[\left(\frac{3V\kappa^3}{4\pi} \right)^{\frac{3-n^{\prime}}{3}}-1 \right]
\bigg\{ -\ln(1-e^{-\frac{\kappa}{T}}) \frac{\kappa^{n+3}}{n+3}\bigg\} +
\frac{u_{n,n^\prime}}{(n+3)} \right ]\nonumber \\
   &+ \displaystyle{\sum_{n=0}^{\infty}} \left[ \frac{1}{\l\pi\r^2}
\frac{a_{n,3}}{n!3! \kappa^{n}} \bigg(\frac{4\pi T}{3V_{ac}\kappa^3}\bigg)
\ln\bigg(\frac{3 \kappa^3 V}{4\pi}\bigg) \bigg\{ -\ln(1-e^{-\frac{\kappa}{T}})
\frac{\kappa^{n+3}}{n+3}\bigg\} + \frac{u_{n,3}}{(n+3)}\right].
 \end{align}
The Helmholtz free energy is,

\begin{align}
 A &=\mu N-PV_{ac} = -PV_{ac}  \nonumber \\
   &= \displaystyle{\sum_{n=0,n^{\prime}=0 \atop n^{\prime}\neq
3}^{\infty}}\left[ \frac{1}{\l\pi\r^2} \frac{a_{n,n^{\prime}}}{n!n^{\prime}!
\kappa^{n+3}} \frac{4\pi T}{\left(3-n^{\prime}\right)}
\left[\left(\frac{3V\kappa^3}{4\pi} \right)^{\frac{3-n^{\prime}}{3}}-1 \right]
\bigg\{ \ln(1-e^{-\frac{\kappa}{T}}) \frac{\kappa^{n+3}}{n+3}\bigg\} -
\frac{u_{n,n^{\prime}} V_{ac}}{(n+3)} \right]  \nonumber \\
   &+\displaystyle{\sum_{n=0}^{\infty}} \frac{a_{n,3}}{n!3! \kappa^{n+3}} \left[
\frac{1}{\l\pi\r^2} \left(\frac{4\pi T}{3}\right) \ln\bigg(\frac{3 \kappa^3
V}{4\pi}\bigg) \bigg\{ \ln(1-e^{-\frac{\kappa}{T}})
\frac{\kappa^{n+3}}{n+3}\bigg\} - \frac{u_{n,3} V_{ac}}{(n+3)} \right].
\end{align}
And the entropy becomes,

\begin{align}\label{modifiedS}
 S=\frac{U-A}{T} &= \displaystyle{\sum_{n=0,n^{\prime}=0 \atop n^{\prime}\neq
3}^{\infty}} \left[ \frac{1}{\l\pi\r^2} \frac{a_{n,n^{\prime}}}{n!n^{\prime}!
\kappa^{n+3}} \frac{4\pi}{\left(3-n^{\prime}\right)}
\left[\left(\frac{3V\kappa^3}{4\pi} \right)^{\frac{3-n^{\prime}}{3}}-1 \right]
\bigg\{ -\ln(1-e^{-\frac{\kappa}{T}}) \frac{\kappa^{n+3}}{n+3}\bigg\} +
\frac{(n+4)}{(n+3)} \left(\frac{u_{n,n^\prime}V_{ac}}{T}\right) \right]
\nonumber \\
   &+\displaystyle{\sum_{n=0}^{\infty}} \left[\frac{1}{\l\pi\r^2}
\frac{a_{n,3}}{n!3! \kappa^{n+3}} \left(\frac{4\pi}{3}\right)
\ln\bigg(\frac{3\kappa^3 V}{4\pi}\bigg) \bigg\{ -\ln(1-e^{-\frac{\kappa}{T}})
\frac{\kappa^{n+3}}{n+3} \bigg\} + \frac{(n+4)}{(n+3)}
\left(\frac{u_{n,3}V_{ac}}{T}\right) \right].
\end{align}
The equilibrium number of photons in modified measure can be estimated in the
same way as we did in unmodified case and using (\ref{modifiedModes}) we have,

\begin{align}\label{modifiedN}
\bar{N} &= \frac{1}{\l\pi\r^2} \displaystyle{\sum_{n=0,n^{\prime}=0 \atop
n^{\prime}\neq 3}^{\infty}} \frac{a_{n,n^{\prime}}}{n!n^{\prime}! k^{n+3}}
\frac{4\pi T^{n+3}}{\left(3-n^{\prime}\right)}
\left[\left(\frac{3V\kappa^3}{4\pi} \right)^{\frac{3-n^{\prime}}{3}}-1 \right]
\Gamma(n+3) \bigg[Z_{n+3}(0)-Z_{n+3}\left(\frac{\kappa}{T}\right)\bigg] 
\nonumber \\
   &+ \frac{1}{\l\pi\r^2}\displaystyle{\sum_{n=0}^{\infty}} \frac{a_{n,3}}{n!3!
\kappa^{n+3}} \left(\frac{4\pi T^{n+3}}{3}\right) \ln\bigg(\frac{3\kappa^3
V}{4\pi}\bigg)  \Gamma(n+3)
\bigg[Z_{n+3}(0)-Z_{n+3}\left(\frac{\kappa}{T}\right)\bigg].
\end{align}

\subsection{The leading high temperature behaviour for unmodified
case}\label{unmodified_limit}
In the high temperature ($T\rightarrow \kappa$) case we get,

\begin{align}
u\approx -\frac{18
\kappa^4}{\pi^2}\left[Z_4(0)-Z_4(1)-\frac{1}{18(e-1)}\right]+\frac{24 \kappa^4
}{\pi^2} \left(\frac{T}{\kappa}\right)
\left[Z_4(0)-Z_4(1)-\frac{1}{24(e-1)}\right],
\end{align}

\begin{align}
P\approx -\frac{6 \kappa^4}{\pi^2} \left[Z_4(0)-Z_4(1)\right]+\frac{8 \kappa^4
}{\pi^2}\left(\frac{T}{\kappa}\right) \left[Z_4(0)-Z_4(1)\right]- \frac{\kappa^4
}{3\pi^2} \left(\frac{T}{\kappa}\right) \ln \left( 1-\frac{1}{e} \right),
\end{align}

\begin{align}
S \approx -\frac{\kappa^3 V_{ac}}{3 \pi^2}\ln\left(1-\frac{1}{e}\right)-\frac{16
\kappa^3 V_{ac}}{\pi^2}\left[Z_4(0)-Z_4(1)-\frac{1}{48(e-1)}\right]+\frac{24
\kappa^3 V_{ac}}{
\pi^2}\left(\frac{T}{\kappa}\right)\left[Z_4(0)-Z_4(1)-\frac{1}{24(e-1)}\right],
\end{align}

\begin{align}
\bar{N} \approx -\frac{4 \kappa^3
V_{ac}}{\pi^2}\left[Z_3(0)-Z_3(1)-\frac{1}{4(e-1)}\right]+\frac{6 \kappa^3
V_{ac}}{\pi^2}\left(\frac{T}{\kappa}\right)
\left[Z_3(0)-Z_3(1)-\frac{1}{6(e-1)}\right]
\end{align}
and
\begin{align}\label{C_V_limit}
C_V \approx -\frac{48 \kappa^3
V_{ac}}{\pi^2}\left[Z_4(0)-Z_4(1)-\frac{(3e-2)}{48(e-1)^2}\right]+\frac{72
\kappa^3 V_{ac} }{\pi^2}\left(\frac{T}{\kappa}\right)
\left[Z_4(0)-Z_4(1)-\frac{(4e-3)}{72(e-1)^2}\right].
\end{align}

\subsection{The leading low temperature behaviour for modified
case}\label{modified_low_limit}
The low temperature limit can be calculated as we did in case of unmodified
measure to get,

\begin{align}
   u &\approx \frac{1}{\l\pi\r^2} \displaystyle{\sum_{n=0,n^{\prime}=0 \atop
n^{\prime}\neq 3}^{\infty}} \frac{a_{n,n^{\prime}}}{n!n^{\prime}! \kappa^{n}}
\frac{4\pi}{\left(3-n^{\prime}\right)}
\bigg(\frac{T^{n+4}}{V_{ac}\kappa^3}\bigg) \left[\left(\frac{3V\kappa^3}{4\pi}
\right)^{\frac{3-n^{\prime}}{3}}-1 \right] \Gamma(n+4) Z_{n+4}(0)  \nonumber \\
   &+ \frac{1}{\l\pi\r^2}\displaystyle{\sum_{n=0}^{\infty}} \frac{a_{n,3}}{n!3!
\kappa^{n}} \bigg(\frac{4\pi T^{n+4}}{3V_{ac}\kappa^3}\bigg) \ln\bigg(\frac{3
\kappa^3 V}{4\pi}\bigg)  \Gamma(n+4) Z_{n+4}(0),
\end{align}

\begin{align}
   P &\approx \frac{1}{\l\pi\r^2} \displaystyle{\sum_{n=0,n^{\prime}=0 \atop
n^{\prime}\neq 3}^{\infty}} \frac{a_{n,n^{\prime}}}{n!n^{\prime}! \kappa^{n}}
\frac{4\pi}{\left(3-n^{\prime}\right)}
\bigg(\frac{T^{n+4}}{V_{ac}\kappa^3}\bigg) \left[\left(\frac{3V\kappa^3}{4\pi}
\right)^{\frac{3-n^{\prime}}{3}}-1 \right] \frac{\Gamma(n+4)}{(n+3)} Z_{n+4}(0) 
\nonumber \\
   &+ \frac{1}{\l\pi\r^2}\displaystyle{\sum_{n=0}^{\infty}} \frac{a_{n,3}}{n!3!
\kappa^{n}} \bigg(\frac{4\pi T^{n+4}}{3V_{ac}\kappa^3}\bigg) \ln\bigg(\frac{3
\kappa^3 V}{4\pi}\bigg)  \frac{\Gamma(n+4)}{(n+3)} Z_{n+4}(0),
\end{align}

\begin{align}
   S &\approx \frac{1}{\l\pi\r^2} \displaystyle{\sum_{n=0,n^{\prime}=0 \atop
n^{\prime}\neq 3}^{\infty}} \frac{a_{n,n^{\prime}}}{n!n^{\prime}! \kappa^{n}}
\frac{4\pi}{\left(3-n^{\prime}\right)} \bigg(\frac{T^{n+3}}{\kappa^3}\bigg)
\left[\left(\frac{3V\kappa^3}{4\pi} \right)^{\frac{3-n^{\prime}}{3}}-1 \right]
\frac{\Gamma(n+4)(n+4)}{(n+3)} Z_{n+4}(0)  \nonumber \\
   &+ \frac{1}{\l\pi\r^2}\displaystyle{\sum_{n=0}^{\infty}} \frac{a_{n,3}}{n!3!
\kappa^{n}} \bigg(\frac{4\pi T^{n+3}}{3\kappa^3}\bigg) \ln\bigg(\frac{3 \kappa^3
V}{4\pi}\bigg)  \frac{\Gamma(n+4)(n+4)}{(n+3)} Z_{n+4}(0),
\end{align}

\begin{align}
   C_V &\approx \frac{1}{\l\pi\r^2} \displaystyle{\sum_{n=0,n^{\prime}=0 \atop
n^{\prime}\neq 3}^{\infty}} \frac{a_{n,n^{\prime}}}{n!n^{\prime}! \kappa^{n}}
\frac{4\pi}{\left(3-n^{\prime}\right)} \bigg(\frac{T^{n+3}}{\kappa^3}\bigg)
\left[\left(\frac{3V\kappa^3}{4\pi} \right)^{\frac{3-n^{\prime}}{3}}-1 \right]
\Gamma(n+4)(n+4) Z_{n+4}(0)  \nonumber \\
   &+ \frac{1}{\l\pi\r^2}\displaystyle{\sum_{n=0}^{\infty}} \frac{a_{n,3}}{n!3!
\kappa^{n}} \bigg(\frac{4\pi T^{n+3}}{3\kappa^3}\bigg) \ln\bigg(\frac{3 \kappa^3
V}{4\pi}\bigg)  \Gamma(n+4)(n+4) Z_{n+4}(0)
\end{align}
and
\begin{align}
\bar{N} &\approx \frac{1}{\l\pi\r^2} \displaystyle{\sum_{n=0,n^{\prime}=0 \atop
n^{\prime}\neq 3}^{\infty}} \frac{a_{n,n^{\prime}}}{n!n^{\prime}! k^{n+3}}
\frac{4\pi T^{n+3}}{\left(3-n^{\prime}\right)}
\left[\left(\frac{3V\kappa^3}{4\pi} \right)^{\frac{3-n^{\prime}}{3}}-1 \right]
\Gamma(n+3) Z_{n+3}(0)  \nonumber \\
   &+ \frac{1}{\l\pi\r^2}\displaystyle{\sum_{n=0}^{\infty}} \frac{a_{n,3}}{n!3!
\kappa^{n+3}} \left(\frac{4\pi T^{n+3}}{3}\right) \ln\bigg(\frac{3\kappa^3
V}{4\pi}\bigg)  \Gamma(n+3) Z_{n+3}(0)
\end{align}

\subsection{The leading high temperature behaviour for modified
case}\label{modified_high_limit}
In high temperature ($T\rightarrow \kappa$) case the behaviour is

\begin{align}
 u \approx u_a+u_b T
\end{align}
where $u_a$ and $u_b$ is

\begin{align}
    u_a &= -\frac{1}{\l\pi\r^2} \displaystyle{\sum_{n=0,n^{\prime}=0 \atop
n^{\prime}\neq 3}^{\infty}} \frac{a_{n,n^{\prime}}}{n!n^{\prime}! }
\frac{4\pi}{\left(3-n^{\prime}\right)}
\bigg(\frac{\kappa^{4}}{V_{ac}\kappa^3}\bigg)
\left[\left(\frac{3V\kappa^3}{4\pi} \right)^{\frac{3-n^{\prime}}{3}}-1 \right]
\Gamma(n+4)(n+3) \bigg[Z_{n+4}(0)-Z_{n+4}(1) - \frac{1}{(n+3)!(n+3)(e-1)} \bigg]
 \nonumber \\
   &- \frac{1}{\l\pi\r^2}\displaystyle{\sum_{n=0}^{\infty}} \frac{a_{n,3}}{n!3!
} \bigg(\frac{4\pi \kappa^{4}}{3V_{ac}\kappa^3}\bigg) \ln\bigg(\frac{3 \kappa^3
V}{4\pi}\bigg)  \Gamma(n+4) (n+3) \bigg[Z_{n+4}(0)-Z_{n+4}(1) - \frac{1}{(n+3)!
(n+3)(e-1)} \bigg]
\end{align}
\begin{align}
    u_b &= \frac{1}{\l\pi\r^2} \displaystyle{\sum_{n=0,n^{\prime}=0 \atop
n^{\prime}\neq 3}^{\infty}} \frac{a_{n,n^{\prime}}}{n!n^{\prime}! }
\frac{4\pi}{\left(3-n^{\prime}\right)}
\bigg(\frac{\kappa^{4}}{V_{ac}\kappa^3}\bigg)
\left[\left(\frac{3V\kappa^3}{4\pi} \right)^{\frac{3-n^{\prime}}{3}}-1 \right]
\Gamma(n+4) \frac{(n+4)}{\kappa} \bigg[Z_{n+4}(0)-Z_{n+4}(1) -
\frac{1}{(n+3)!(n+4)(e-1)} \bigg]  \nonumber \\
   &+ \frac{1}{\l\pi\r^2}\displaystyle{\sum_{n=0}^{\infty}} \frac{a_{n,3}}{n!3!
} \bigg(\frac{4\pi \kappa^{4}}{3V_{ac}\kappa^3}\bigg) \ln\bigg(\frac{3 \kappa^3
V}{4\pi}\bigg)  \Gamma(n+4) \frac{(n+4)}{\kappa} \bigg[Z_{n+4}(0)-Z_{n+4}(1) -
\frac{1}{(n+3)! (n+4)(e-1)} \bigg],
\end{align}
\begin{align}
 P \approx P_a+P_b T
\end{align}
where $P_a$ and $P_b$ is

\begin{align}
    P_a &= -\frac{1}{\l\pi\r^2} \displaystyle{\sum_{n=0,n^{\prime}=0 \atop
n^{\prime}\neq 3}^{\infty}} \frac{a_{n,n^{\prime}}}{n!n^{\prime}! }
\frac{4\pi}{\left(3-n^{\prime}\right)}
\bigg(\frac{\kappa^{4}}{V_{ac}\kappa^3}\bigg)
\left[\left(\frac{3V\kappa^3}{4\pi} \right)^{\frac{3-n^{\prime}}{3}}-1
\right]\Gamma(n+4) \bigg[Z_{n+4}(0)-Z_{n+4}(1)\bigg]  \nonumber \\
   &- \frac{1}{\l\pi\r^2}\displaystyle{\sum_{n=0}^{\infty}} \frac{a_{n,3}}{n!3!
} \bigg(\frac{4\pi \kappa^{4}}{3V_{ac}\kappa^3}\bigg) \ln\bigg(\frac{3 \kappa^3
V}{4\pi}\bigg)\Gamma(n+4) \bigg[Z_{n+4}(0)-Z_{n+4}(1)\bigg]  
\end{align}
\begin{align}
    P_b &= \frac{1}{\l\pi\r^2} \displaystyle{\sum_{n=0,n^{\prime}=0 \atop
n^{\prime}\neq 3}^{\infty}} \frac{a_{n,n^{\prime}}}{n!n^{\prime}! }
\frac{4\pi}{\left(3-n^{\prime}\right)}
\bigg(\frac{\kappa^{4}}{V_{ac}\kappa^3}\bigg)
\left[\left(\frac{3V\kappa^3}{4\pi} \right)^{\frac{3-n^{\prime}}{3}}-1 \right]
\left\{ \Gamma(n+4)
\frac{(n+4)}{\kappa(n+3)}\bigg[Z_{n+4}(0)-Z_{n+4}(1)\bigg]-\frac{1}{\kappa(n+3)}
\ln \left(1-\frac{1}{e}\right) \right\}  \nonumber \\
   &+ \frac{1}{\l\pi\r^2}\displaystyle{\sum_{n=0}^{\infty}} \frac{a_{n,3}}{n!3!
} \bigg(\frac{4\pi \kappa^{4}}{3V_{ac}\kappa^3}\bigg) \ln\bigg(\frac{3 \kappa^3
V}{4\pi}\bigg)  \left\{ \Gamma(n+4)\frac{(n+4)}{\kappa(n+3)}
\bigg[Z_{n+4}(0)-Z_{n+4}(1)\bigg]-\frac{1}{\kappa(n+3)}\ln
\left(1-\frac{1}{e}\right) \right\},  
\end{align}
\begin{align}
 S \approx S_a+S_b T
\end{align}
where $S_a$ and $S_b$ is

\begin{align}
    S_a &= -\frac{1}{\l\pi\r^2} \displaystyle{\sum_{n=0,n^{\prime}=0 \atop
n^{\prime}\neq 3}^{\infty}} \frac{a_{n,n^{\prime}}}{n!n^{\prime}! }
\frac{4\pi}{\left(3-n^{\prime}\right)} \left[\left(\frac{3V\kappa^3}{4\pi}
\right)^{\frac{3-n^{\prime}}{3}}-1 \right] \frac{1}{(n+3)}
\bigg\{\Gamma(n+4)(n+4)(n+2) \bigg[Z_{n+4}(0)-Z_{n+4}(1) \nonumber \\ 
      &-\frac{1}{(n+3)!(n+2)(e-1)} \bigg]+\frac{1}{(e-1)}+\ln
\left(1-\frac{1}{e}\right) \bigg \} \nonumber \\ 
     &- \frac{1}{\l\pi\r^2}\displaystyle{\sum_{n=0}^{\infty}}
\frac{a_{n,3}}{n!3! } \bigg(\frac{4\pi}{3}\bigg) \ln\bigg(\frac{3 \kappa^3
V}{4\pi}\bigg) \frac{1}{(n+3)} \bigg\{\Gamma(n+4)(n+4)(n+2)
\bigg[Z_{n+4}(0)-Z_{n+4}(1) \nonumber \\ 
     &-\frac{1}{(n+3)!(n+2)(e-1)} \bigg]+\frac{1}{(e-1)}+\ln
\left(1-\frac{1}{e}\right)\bigg\}
\end{align}
\begin{align}
    S_b &= -\frac{1}{\l\pi\r^2} \displaystyle{\sum_{n=0,n^{\prime}=0 \atop
n^{\prime}\neq 3}^{\infty}} \frac{a_{n,n^{\prime}}}{n!n^{\prime}! }
\frac{4\pi}{\left(3-n^{\prime}\right)} \left[\left(\frac{3V\kappa^3}{4\pi}
\right)^{\frac{3-n^{\prime}}{3}}-1 \right] 
\bigg\{\Gamma(n+4)\frac{(n+4)}{\kappa} \bigg[Z_{n+4}(0)-Z_{n+4}(1) \nonumber \\ 
      &-\frac{1}{(n+3)!(n+4)(e-1)} \bigg] \bigg \} \nonumber \\ 
     &- \frac{1}{\l\pi\r^2}\displaystyle{\sum_{n=0}^{\infty}}
\frac{a_{n,3}}{n!3! } \bigg(\frac{4\pi}{3}\bigg) \ln\bigg(\frac{3 \kappa^3
V}{4\pi}\bigg) \bigg\{\Gamma(n+4)\frac{(n+4)}{\kappa}
\bigg[Z_{n+4}(0)-Z_{n+4}(1) \nonumber \\ 
     &-\frac{1}{(n+3)!(n+4)(e-1)} \bigg] \bigg\},
\end{align}
\begin{align}
 C_V \approx {C_V}_a+{C_V}_b T
\end{align}
where ${C_V}_a$ and ${C_V}_b$ is

\begin{align}
    {C_V}_a &= -\frac{1}{\l\pi\r^2} \displaystyle{\sum_{n=0,n^{\prime}=0 \atop
n^{\prime}\neq 3}^{\infty}} \frac{a_{n,n^{\prime}}}{n!n^{\prime}! }
\frac{4\pi}{\left(3-n^{\prime}\right)} \left[\left(\frac{3V\kappa^3}{4\pi}
\right)^{\frac{3-n^{\prime}}{3}}-1 \right] \bigg\{ \Gamma(n+4)(n+4)(n+2)
\bigg[Z_{n+4}(0)-Z_{n+4}(1) \nonumber \\ 
    &-\frac{1}{(n+3)!(n+2)(e-1)} \bigg]+\frac{(e-2)}{(e-1)^2}\bigg\} \nonumber
\\ 
   &- \frac{1}{\l\pi\r^2}\displaystyle{\sum_{n=0}^{\infty}} \frac{a_{n,3}}{n!3!
} \bigg(\frac{4\pi}{3}\bigg) \ln\bigg(\frac{3 \kappa^3 V}{4\pi}\bigg) \bigg\{
\Gamma(n+4)(n+4)(n+2) \bigg[Z_{n+4}(0)-Z_{n+4}(1) \nonumber \\ 
   &-\frac{1}{(n+3)!(n+2)(e-1)} \bigg]+\frac{(e-2)}{(e-1)^2}\bigg\}
    \end{align}
\begin{align}
    {C_V}_b &= \frac{1}{\l\pi\r^2} \displaystyle{\sum_{n=0,n^{\prime}=0 \atop
n^{\prime}\neq 3}^{\infty}} \frac{a_{n,n^{\prime}}}{n!n^{\prime}! }
\frac{4\pi}{\left(3-n^{\prime}\right)} \left[\left(\frac{3V\kappa^3}{4\pi}
\right)^{\frac{3-n^{\prime}}{3}}-1 \right] \bigg\{
\Gamma(n+4)(n+4)(n+3)\frac{1}{\kappa}
\bigg[Z_{n+4}(0)-Z_{n+4}(1)-\frac{1}{(n+3)!(n+3)(e-1)}   \nonumber \\
   &+\frac{1}{(n+3)!(n+4)(n+3)(e-1)^2} \bigg] \bigg\} \nonumber \\
   &+ \frac{1}{\l\pi\r^2}\displaystyle{\sum_{n=0}^{\infty}} \frac{a_{n,3}}{n!3!
} \bigg(\frac{4\pi}{3}\bigg) \ln\bigg(\frac{3 \kappa^3 V}{4\pi}\bigg) \bigg\{
\Gamma(n+4)(n+4)(n+3)\frac{1}{\kappa}
\bigg[Z_{n+4}(0)-Z_{n+4}(1)-\frac{1}{(n+3)!(n+3)(e-1)}   \nonumber \\
   &+\frac{1}{(n+3)!(n+4)(n+3)(e-1)^2} \bigg] \bigg\}
    \end{align}
and    
\begin{align}
 \bar{N} \approx {\bar{N}}_a+{\bar{N}}_b T
\end{align}
where ${\bar{N}}_a$ and ${\bar{N}}_b$ is

\begin{align}
 {\bar{N}}_a &= -\frac{1}{\l\pi\r^2} \displaystyle{\sum_{n=0,n^{\prime}=0 \atop
n^{\prime}\neq 3}^{\infty}} \frac{a_{n,n^{\prime}}}{n!n^{\prime}!}
\frac{4\pi}{\left(3-n^{\prime}\right)} \left[\left(\frac{3V\kappa^3}{4\pi}
\right)^{\frac{3-n^{\prime}}{3}}-1 \right] \Gamma(n+3) \bigg \{(n+2)
\bigg[Z_{n+3}(0)-Z_{n+3}(1)-\frac{1}{(n+2)!(n+2)(e-1)}\bigg]\bigg\} \nonumber \\
 &- \frac{1}{\l\pi\r^2}\displaystyle{\sum_{n=0}^{\infty}} \frac{a_{n,3}}{n!3!}
\left(\frac{4\pi }{3}\right) \ln\bigg(\frac{3\kappa^3 V}{4\pi}\bigg) 
\Gamma(n+3) \bigg \{(n+2)
\bigg[Z_{n+3}(0)-Z_{n+3}(1)-\frac{1}{(n+2)!(n+2)(e-1)}\bigg]\bigg\}
\end{align}
\begin{align}
 {\bar{N}}_b &= \frac{1}{\l\pi\r^2} \displaystyle{\sum_{n=0,n^{\prime}=0 \atop
n^{\prime}\neq 3}^{\infty}} \frac{a_{n,n^{\prime}}}{n!n^{\prime}!}
\frac{4\pi}{\left(3-n^{\prime}\right)} \left[\left(\frac{3V\kappa^3}{4\pi}
\right)^{\frac{3-n^{\prime}}{3}}-1 \right] \Gamma(n+3) \bigg \{
\frac{(n+3)}{\kappa}
\bigg[Z_{n+3}(0)-Z_{n+3}(1)-\frac{1}{(n+2)!(n+3)(e-1)}\bigg]\bigg\} \nonumber \\
 &+ \frac{1}{\l\pi\r^2}\displaystyle{\sum_{n=0}^{\infty}} \frac{a_{n,3}}{n!3!}
\left(\frac{4\pi }{3}\right) \ln\bigg(\frac{3\kappa^3 V}{4\pi}\bigg) 
\Gamma(n+3) \bigg \{\frac{(n+3)}{\kappa}
\bigg[Z_{n+3}(0)-Z_{n+3}(1)-\frac{1}{(n+2)!(n+3)(e-1)}\bigg]\bigg\}
\end{align}
\end{appendices}


\begin{thebibliography}{999}


\bibitem{AmelinoCamelia:1997gz} 
  G.~Amelino-Camelia, J.~R.~Ellis, N.~E.~Mavromatos, D.~V.~Nanopoulos and
S.~Sarkar,
  Nature {\bf 393}, 763 (1998)
  doi:10.1038/31647
  [astro-ph/9712103].
\bibitem{AmelinoCamelia:2000mn} 
  G.~Amelino-Camelia,
  Int.\ J.\ Mod.\ Phys.\ D {\bf 11}, 35 (2002)
  doi:10.1142/S0218271802001330
  [gr-qc/0012051].
  
\bibitem{AmelinoCamelia:2000ge} 
  G.~Amelino-Camelia,
  Phys.\ Lett.\ B {\bf 510}, 255 (2001)
  doi:10.1016/S0370-2693(01)00506-8
  [hep-th/0012238].

\bibitem{AmelinoCamelia:1999pm} 
  G.~Amelino-Camelia and S.~Majid,
  Int.\ J.\ Mod.\ Phys.\ A {\bf 15}, 4301 (2000)
  doi:10.1142/S0217751X00002777, 10.1142/S0217751X00002779
  [hep-th/9907110].
  
\bibitem{Jacobson:2005bg} 
  T.~Jacobson, S.~Liberati and D.~Mattingly,
  Annals Phys.\  {\bf 321}, 150 (2006)
  doi:10.1016/j.aop.2005.06.004
  [astro-ph/0505267].
 
\bibitem{Shao:2009bv} 
  L.~Shao, Z.~Xiao and B.~Q.~Ma,
  Astropart.\ Phys.\  {\bf 33}, 312 (2010)
  doi:10.1016/j.astropartphys.2010.03.003
  [arXiv:0911.2276 [hep-ph]].
  
\bibitem{Shao:2010wk} 
  L.~Shao and B.~Q.~Ma,
  Mod.\ Phys.\ Lett.\ A {\bf 25}, 3251 (2010)
  doi:10.1142/S0217732310034572
  [arXiv:1007.2269 [hep-ph]].

  
\bibitem{Magueijo:2001cr} 
  J.~Magueijo and L.~Smolin,
  Phys.\ Rev.\ Lett.\  {\bf 88}, 190403 (2002)
  doi:10.1103/PhysRevLett.88.190403
  [hep-th/0112090].
  
\bibitem{Magueijo:2002am}
  J.~Magueijo, L.~Smolin,
  Phys.\ Rev.\  {\bf D67}, 044017 (2003).
  [gr-qc/0207085].
 
\bibitem{Chandra:2011nj} 
  N.~Chandra and S.~Chatterjee,
  Phys.\ Rev.\ D {\bf 85}, 045012 (2012)
  doi:10.1103/PhysRevD.85.045012
  [arXiv:1108.0896 [gr-qc]].
 
\bibitem{KowalskiGlikman:2002we} 
  J.~Kowalski-Glikman and S.~Nowak,
  Phys.\ Lett.\ B {\bf 539}, 126 (2002)
  doi:10.1016/S0370-2693(02)02063-4
  [hep-th/0203040].

\bibitem{Snyder:1946qz} 
  H.~S.~Snyder,
  Phys.\ Rev.\  {\bf 71}, 38 (1947).
  doi:10.1103/PhysRev.71.38
  
\bibitem{Chandra:2014qva} 
  N.~Chandra, H.~W.~Groenewald, J.~N.~Kriel, F.~G.~Scholtz and S.~Vaidya,
  J.\ Phys.\ A {\bf 47}, no. 44, 445203 (2014)
  doi:10.1088/1751-8113/47/44/445203
  [arXiv:1407.5857 [hep-th]].
  
\bibitem{Gross:1987ar} 
  D.~J.~Gross and P.~F.~Mende,
  Nucl.\ Phys.\ B {\bf 303}, 407 (1988).
  doi:10.1016/0550-3213(88)90390-2
  
\bibitem{Amati:1988tn} 
  D.~Amati, M.~Ciafaloni and G.~Veneziano,
  Phys.\ Lett.\ B {\bf 216}, 41 (1989).
  doi:10.1016/0370-2693(89)91366-X
  
\bibitem{Maggiore:1993kv} 
  M.~Maggiore,
  Phys.\ Lett.\ B {\bf 319}, 83 (1993)
  doi:10.1016/0370-2693(93)90785-G
  [hep-th/9309034].
  
\bibitem{Garay:1994en} 
  L.~J.~Garay,
  Int.\ J.\ Mod.\ Phys.\ A {\bf 10}, 145 (1995)
  doi:10.1142/S0217751X95000085
  [gr-qc/9403008].

\bibitem{Kempf:1994su} 
  A.~Kempf, G.~Mangano and R.~B.~Mann,
  Phys.\ Rev.\ D {\bf 52}, 1108 (1995)
  doi:10.1103/PhysRevD.52.1108
  [hep-th/9412167].

\bibitem{Kempf:1996nk} 
  A.~Kempf and G.~Mangano,
  Phys.\ Rev.\ D {\bf 55}, 7909 (1997)
  doi:10.1103/PhysRevD.55.7909
  [hep-th/9612084].
   
  
\bibitem{Smailagic:2003yb} 
  A.~Smailagic and E.~Spallucci,
  J.\ Phys.\ A {\bf 36}, L467 (2003)
  doi:10.1088/0305-4470/36/33/101
  [hep-th/0307217].
  
\bibitem{Smailagic:2003rp} 
  A.~Smailagic and E.~Spallucci,
  J.\ Phys.\ A {\bf 36}, L517 (2003)
  doi:10.1088/0305-4470/36/39/103
  [hep-th/0308193].

  
\bibitem{Kober:2010um} 
  M.~Kober and P.~Nicolini,
  Class.\ Quant.\ Grav.\  {\bf 27}, 245024 (2010)
  doi:10.1088/0264-9381/27/24/245024
  [arXiv:1005.3293 [hep-th]].
  
\bibitem{Nozari:2012gd} 
  K.~Nozari and A.~Etemadi,
  Phys.\ Rev.\ D {\bf 85}, 104029 (2012)
  doi:10.1103/PhysRevD.85.104029
  [arXiv:1205.0158 [hep-th]].
  
  
 
\bibitem{Magueijo:2002xx} 
  J.~Magueijo and L.~Smolin,
  Class.\ Quant.\ Grav.\  {\bf 21}, 1725 (2004)
  doi:10.1088/0264-9381/21/7/001
  [gr-qc/0305055].
  
\bibitem{Smolin:2005cz} 
  L.~Smolin,
  Nucl.\ Phys.\ B {\bf 742}, 142 (2006)
  doi:10.1016/j.nuclphysb.2006.02.017
  [hep-th/0501091].
  
\bibitem{Gorji:2016laj} 
  M.~A.~Gorji, K.~Nozari and B.~Vakili,
  Phys.\ Lett.\ B {\bf 765}, 113 (2017)
  doi:10.1016/j.physletb.2016.12.023
  [arXiv:1606.00910 [gr-qc]].
  
\bibitem{Ali:2014zea} 
  A.~F.~Ali, M.~Faizal and M.~M.~Khalil,
  Nucl.\ Phys.\ B {\bf 894}, 341 (2015)
  doi:10.1016/j.nuclphysb.2015.03.014
  [arXiv:1410.5706 [hep-th]].
  
\bibitem{Ling:2005bp} 
  Y.~Ling, X.~Li and H.~b.~Zhang,
  Mod.\ Phys.\ Lett.\ A {\bf 22}, 2749 (2007)
  doi:10.1142/S0217732307022931
  [gr-qc/0512084].
  
  
\bibitem{KowalskiGlikman:2001ct} 
  J.~Kowalski-Glikman,
  Phys.\ Lett.\ A {\bf 299}, 454 (2002)
  doi:10.1016/S0375-9601(02)00751-X
  [hep-th/0111110].
  
\bibitem{Blaut:2003wg} 
  A.~Blaut, M.~Daszkiewicz, J.~Kowalski-Glikman and S.~Nowak,
  Phys.\ Lett.\ B {\bf 582}, 82 (2004)
  doi:10.1016/j.physletb.2003.12.035
  [hep-th/0312045].
  
\bibitem{Borowiec:2009vb} 
  A.~Borowiec and A.~Pachol,
  J.\ Phys.\ A {\bf 43}, 045203 (2010)
  doi:10.1088/1751-8113/43/4/045203
  [arXiv:0903.5251 [hep-th]].
  
\bibitem{Magpantay:2010zz} 
  J.~A.~Magpantay,
  Int.\ J.\ Mod.\ Phys.\ A {\bf 25}, 1881 (2010)
  doi:10.1142/S0217751X1004807X
  [arXiv:1011.3662 [math-ph]].
  
\bibitem{Pramanik:2012fj} 
  S.~Pramanik, S.~Ghosh and P.~Pal,
  Annals Phys.\  {\bf 346}, 113 (2014)
  doi:10.1016/j.aop.2014.04.009
  [arXiv:1212.6881 [hep-th]].
  
\bibitem{Camacho:2007qy} 
  A.~Camacho and A.~Macias,
  Gen.\ Rel.\ Grav.\  {\bf 39}, 1175 (2007)
  doi:10.1007/s10714-007-0419-1
  [gr-qc/0702150 [GR-QC]].

\bibitem{Zhang:2011ms} 
  X.~Zhang, L.~Shao and B.~Q.~Ma,
  Astropart.\ Phys.\  {\bf 34}, 840 (2011)
  doi:10.1016/j.astropartphys.2011.03.001
  [arXiv:1102.2613 [hep-th]].

\bibitem{Grether:2007ur} 
  M.~Grether, M.~de Llano and G.~A.~Baker, Jr,
  Phys.\ Rev.\ Lett.\  {\bf 99}, 200406 (2007)
  doi:10.1103/PhysRevLett.99.200406
  [arXiv:0706.2833 [cond-mat.supr-con]].

\bibitem{Gregg:2008jb} 
  M.~Gregg and S.~A.~Major,
  Int.\ J.\ Mod.\ Phys.\ D {\bf 18}, 971 (2009)
  doi:10.1142/S021827180901487X
  [arXiv:0806.3496 [astro-ph]].
  
\bibitem{Moussa:2015yqy} 
  M.~Moussa,
  Physica {\bf 465}, 25 (2017)
  doi:10.1016/j.physa.2016.08.005
  [arXiv:1511.06183 [physics.gen-ph]].
  
  
\bibitem{Das:2010gk} 
  S.~Das and D.~Roychowdhury,
  Phys.\ Rev.\ D {\bf 81}, 085039 (2010)
  doi:10.1103/PhysRevD.81.085039
  [arXiv:1002.0192 [hep-th]].
  
  \bibitem{pathria} R K Pathria, {\it Statistical Mechanics, Academic Press Inc;
3rd Revised edition edition}.

\bibitem{Faruk:2016brl} 
  M.~M.~Faruk and M.~M.~Rahman,
  arXiv:1605.06784v2 [hep-th].
  
\bibitem{Scholtz:2015fba} 
  F.~G.~Scholtz, J.~N.~Kriel and H.~W.~Groenewald,
  Phys.\ Rev.\ D {\bf 92}, no. 12, 125013 (2015)
  doi:10.1103/PhysRevD.92.125013
  [arXiv:1508.05799 [hep-th]].
  
\bibitem{Corichi:2007tf} 
  A.~Corichi, T.~Vukasinac and J.~A.~Zapata,
  Phys.\ Rev.\ D {\bf 76}, 044016 (2007)
  doi:10.1103/PhysRevD.76.044016
  [arXiv:0704.0007 [gr-qc]].
 
\bibitem{Hossain:2010wy} 
  G.~M.~Hossain, V.~Husain and S.~S.~Seahra,
  Class.\ Quant.\ Grav.\  {\bf 27}, 165013 (2010)
  doi:10.1088/0264-9381/27/16/165013
  [arXiv:1003.2207 [gr-qc]].
  
  
\bibitem{Corichi:2012bg} 
  A.~Corichi and T.~Vukasinac,
  Phys.\ Rev.\ D {\bf 86}, 064019 (2012)
  doi:10.1103/PhysRevD.86.064019
  [arXiv:1202.1846 [gr-qc]].
 
\bibitem{Majumder:2012qy} 
  B.~Majumder and S.~Sen,
  Phys.\ Lett.\ B {\bf 717}, 291 (2012)
  doi:10.1016/j.physletb.2012.09.035
  [arXiv:1207.6459 [gr-qc]].
  

\bibitem{Liberati:2004ju} 
  S.~Liberati, S.~Sonego and M.~Visser,
  Phys.\ Rev.\ D {\bf 71}, 045001 (2005)
  doi:10.1103/PhysRevD.71.045001
  [gr-qc/0410113].
  
\bibitem{Hossenfelder:2007fy} 
  S.~Hossenfelder,
  Phys.\ Rev.\ D {\bf 75}, 105005 (2007)
  doi:10.1103/PhysRevD.75.105005
  [hep-th/0702016].
  
\bibitem{Hossenfelder:2014ifa} 
  S.~Hossenfelder,
  SIGMA {\bf 10}, 074 (2014)
  doi:10.3842/SIGMA.2014.074
  [arXiv:1403.2080 [gr-qc]].
  
\bibitem{Mandanici:2007eb} 
  G.~Mandanici,
  Mod.\ Phys.\ Lett.\ A {\bf 24}, 739 (2009)
  doi:10.1142/S0217732309030424
  [arXiv:0707.3700 [gr-qc]].

\bibitem{Deriglazov:2004hg} 
  A.~A.~Deriglazov and B.~F.~Rizzuti,
  Phys.\ Rev.\ D {\bf 71}, 123515 (2005)
  doi:10.1103/PhysRevD.71.123515
  [hep-th/0410087].
  
\bibitem{Girelli:2004ue} 
  F.~Girelli and E.~R.~Livine,
  Braz.\ J.\ Phys. {\bf 35(2b)}, 432-438 (2005)
  doi:10.1590/S0103-97332005000300011
  [gr-qc/0412004].
  
\bibitem{Deriglazov:2004yr} 
  A.~A.~Deriglazov,
  Phys.\ Lett.\ B {\bf 603}, 124 (2004)
  doi:10.1016/j.physletb.2004.10.024
  [hep-th/0409232].
 
 
\bibitem{abramowitz} Milton Abramowitz and Irene A. Stegun, {\it Handbook of
Mathematical Functions, Dover Publications Inc}. 
\bibitem{ashcroft} N.W.Ashcroft, N.D.Mermin, {\it Solid State Physics},
Brooks/Cole.
\bibitem{debye} Shubin, Mikhail, Sunada, Toshikazu, Pure and Appl. Math.
Quarterly 2, 745-777, 2006.
\bibitem{polylogarithm} Wood, D.C. (June 1992). The Computation of
Polylogarithms. Technical Report 15-92.

\bibitem{AmelinoCamelia:2003xp} 
  G.~Amelino-Camelia, L.~Smolin and A.~Starodubtsev,
  Class.\ Quant.\ Grav.\  {\bf 21}, 3095 (2004)
  doi:10.1088/0264-9381/21/13/002
  [hep-th/0306134].
  
\bibitem{Assaniousssi:2014ota} 
  M.~Assanioussi, A.~Dapor and J.~Lewandowski,
  Phys.\ Lett.\ B {\bf 751}, 302 (2015)
  doi:10.1016/j.physletb.2015.10.043
  [arXiv:1412.6000 [gr-qc]].
  
\bibitem{Loret:2015iia} 
  N.~Loret and L.~Barcaroli,
  doi:10.1142/9789814623995-0141, 10.1142/9789814623995-0402
  arXiv:1501.03698 [gr-qc].
  
\bibitem{Assanioussi:2016yxx} 
  M.~Assanioussi and A.~Dapor,
  Phys.\ Rev.\ D {\bf 95}, no. 6, 063513 (2017)
  doi:10.1103/PhysRevD.95.063513
  [arXiv:1606.09186 [gr-qc]].

\bibitem{carroll} Sean M. Carroll, {\it Spacetime and Geometry: An Introduction
to General Relativity, Pearson}.

\bibitem{Ashtekar:2008ay} 
  A.~Ashtekar,
  J.\ Phys.\ Conf.\ Ser.\  {\bf 189}, 012003 (2009)
  doi:10.1088/1742-6596/189/1/012003
  [arXiv:0812.4703 [gr-qc]].
  
\bibitem{Bojowald:2010xq} 
  M.~Bojowald,
  arXiv:1002.2618 [gr-qc].
  
\bibitem{Alesci:2016xqa} 
  E.~Alesci, G.~Botta, F.~Cianfrani and S.~Liberati,
  arXiv:1612.07116 [gr-qc].
  
  
\bibitem{WhiteD} D.K. Mishra and N. Chandra, Quantum Gravity effects in White
Dwarfs (under preparation).
\bibitem{rudin} Walter Rudin, {\it Real and Complex Analysis, Third Edition,
McGraw-Hill Book Company}.  


\end{thebibliography}
\end{document}